\documentclass[11pt]{article}
\pdfoutput=1
\usepackage[utf8x]{inputenc}	%for german Umlaute
\usepackage[english]{babel}
\usepackage{amssymb}
\usepackage{amsthm}
\usepackage{mathtools}		%for definition ":="
\usepackage{dsfont}		%for math modes
\usepackage{stmaryrd}		%GIT quotient
\usepackage{cite}		%for better citing
\usepackage[svgnames]{xcolor}
\usepackage{enumerate}
\usepackage{setspace}
\usepackage{booktabs}
\usepackage{tikz}
\usetikzlibrary{shapes,snakes}
\usetikzlibrary{positioning}
\usetikzlibrary{chains}
\usetikzlibrary{arrows,fit,decorations.pathreplacing, shapes.misc}
\tikzstyle{every picture}+=[remember picture]
\tikzstyle{na} = [baseline=-.5ex]
\usepackage{todonotes}
\usepackage[font={small,it},labelfont=bf]{caption} % make captions ``small'' and 
\usepackage{subcaption}
\usepackage[pdftex,breaklinks,colorlinks=true,urlcolor=RoyalBlue,linkcolor=blue,
citecolor=blue]{hyperref}		%for links on refs etc.
\catcode`@=11
\renewcommand{\section}{\@startsection{section}{1}{0pt}{\medskipamount}
{\medskipamount}{\Large\bf}}
\numberwithin{equation}{section}
\catcode`@=12

\newcommand{\im}{\mathrm{i}}
\def\for{\qquad\textrm{for}\quad}
\def\with{\qquad\textrm{with}\quad}
\def\und{\quad\textrm{and}\quad}
\def\st{\quad\textrm{such that}\quad}
\def\rank{\mathrm{rk}}

\def\deg{\mathrm{deg}}
\def\dim{\mathrm{dim}}
\def\codim{\mathrm{codim}}

\newcommand{\HS}{\mathrm{HS}}
\newcommand{\HilbS}{\mathrm{H}}
\newcommand{\PoinS}{\mathrm{P}}
\newcommand{\Hilb}{\mathrm{Hilb}}
\newcommand{\PL}{\mathrm{PL}}
\newcommand{\Relint}{\mathrm{Relint}}
\newcommand{\C}{\mathbb C}
\newcommand{\R}{\mathbb R}
\newcommand{\K}{\mathbb K}
\newcommand{\NN}{\mathbb N}  
\newcommand{\Z}{\mathbb Z}

\newcommand{\MCoulomb}{\mathcal{M}_C}
\newcommand{\MHiggs}{\mathcal{M}_H}
\newcommand{\Ncal}{\mathcal{N}}

\newcommand{\spanN}{\mathrm{Span}_{\NN}}
\newcommand{\cone}{\mathrm{Cone}}
\newcommand{\para}[2]{\mathrm{par}_{#1}(#2)}
\newcommand{\paraprim}[2]{\mathrm{par'}_{#1}(#2)}
\newcommand{\bt}{\boldsymbol{t}}
\newcommand{\bz}{\boldsymbol{z}}
\newcommand{\ba}{\boldsymbol{\alpha}}
\newcommand{\bg}{\boldsymbol{\gamma}}

\newcommand{\Kern}{\mathrm{Ker}}

\newcommand{\gfrak}{\mathfrak{g}}  
\newcommand{\hfrak}{\mathfrak{h}}
\newcommand{\tfrak}{\mathfrak{t}}
  
\newcommand{\Wcal}{\mathcal{W}}
\newcommand{\su}{{{\rm SU}(2)}}

\newcommand{\uo}{{{\rm U}(1)}}

\newcommand{\surm}{{{\rm SU}}}

\newcommand{\sorm}{{{\rm SO}}}
\newcommand{\orm}{{{\rm O}}}

\newcommand{\sprm}{{{\rm Sp}}}

\newcommand{\G}{\mathrm{G}}
\newcommand{\Hh}{\mathrm{H}}
\newcommand{\T}{\mathrm{T}}
\newcommand{\GNOG}{\widehat{\G}}

\newcommand{\Fcal}{\mathcal{F}}
\newcommand{\Kcal}{\mathcal{K}}

\newcommand{\Poincare}{Poincar\'{e}}

\newtheorem{myLemma}{Lemma}
\newtheorem{myProp}{Proposition}
\newtheorem{myConj}{Conjecture}
\newtheorem{myEx}{Example}
\allowdisplaybreaks

\begin{document}
\begin{titlepage}
\setcounter{page}{0}
%
% 
% just a tool to get the last revision date
% \begin{flushleft}
%  \textsc{---DRAFT---}\\
%  \textsc{last compiled: \today}
% \end{flushleft}

\begin{flushright}
Imperial/TP/16/AH/06\\
UWTHPH-2016-23 
\end{flushright}

\vskip 2cm

\begin{center}

{\Large\bf Algebraic properties of the monopole formula
}

\vspace{15mm}

{\large Amihay Hanany${}^{1}$} , \ {\large Marcus Sperling${}^{2}$} 
\\[5mm]
\noindent ${}^1${\em Theoretical Physics Group, Imperial College London\\
Prince Consort Road, London, SW7 2AZ, UK}\\
{Email: {\tt a.hanany@imperial.ac.uk}}
\\[5mm]
\noindent ${}^{2}${\em Fakultät für Physik, Universität Wien}\\
{\em Boltzmanngasse 5, 1200 Wien, Austria}\\
Email: {\tt marcus.sperling@univie.ac.at}
\\[5mm]

\vspace{15mm}

\begin{abstract}
The monopole formula provides the Hilbert series of the Coulomb branch for a 
$3$-dimensional $\Ncal=4$ gauge theory. Employing the concept of a fan defined 
by the matter content, and summing over the corresponding collection of 
monoids, allows the following: 
firstly, we provide explicit expressions for the Hilbert series for any gauge 
group. Secondly, we prove that the order of the pole at $t=1$ and 
$t\to \infty$ equals the complex or quaternionic dimension of the moduli space, 
respectively. Thirdly, we determine all bare and dressed BPS monopole operators 
that are sufficient to generate the entire chiral ring. 
As an application, we demonstrate the implementation of our approach to 
computer algebra programs and the applicability to higher rank gauge theories.
\end{abstract}

\end{center}

\end{titlepage}

{\baselineskip=12pt
{\footnotesize
\tableofcontents
}
}

%%%%%%%%%%%%%%%%%%%%%%%%%%%%%%%%%%%%%%%%%%%%%%%%%%%%%%%%%%%%%%%%%%%%%%%%%%%%%%%%
  \section{Introduction}
\label{sec:introduction}
The moduli spaces of $3$-dimensional supersymmetric gauge 
theories with $8$ supercharges have revealed various interesting features. The 
two prominent branches, Coulomb $\MCoulomb$ and Higgs $\MHiggs$, lie both in 
the family of hyper-Kähler spaces, but behave fundamentally different as 
exhibited in their dimension and behaviour under quantum-corrections, for 
instance.

There have been various attempts to understand the Coulomb branch from a 
variety of perspectives.
Here, we focus on the viewpoint introduced in~\cite{Cremonesi:2013lqa}, which 
introduced the \emph{monopole formula} as a prescription of the Hilbert series 
for the chiral ring $\C[\MCoulomb]$. This approach has been applied to 
various 
questions~\cite{Cremonesi:2014vla,Cremonesi:2014kwa,Cremonesi:2014xha,
Cremonesi:2014uva, Hanany:2016ezz}, and found extensions to $\Ncal\geq2$ 
theories~\cite{Hanany:2015via,Cremonesi:2015dja,Cremonesi:2016nbo} as well as mixed 
branches~\cite{Carta:2016fjb}.
Other approaches attempt a mathematically rigorous definition of 
the Coulomb branch~\cite{Nakajima:2015txa,Nakajima:2015gxa,Braverman:2016wma} or 
study the quantised chiral ring~\cite{Bullimore:2015lsa,Bullimore:2016hdc}.

\paragraph{Monopole formula}
We recall the monopole formula for a $3$-dimensional $\Ncal=4$ gauge theory 
with gauge group $\G$ as
\begin{equation}
 \HS_{\G}(t) = \sum_{m \in \Lambda_w(\GNOG) \slash \Wcal_{\GNOG}} t^{\Delta(m)} 
P_{\G}(t;m) \; , \label{eqn:def_monopole_formula}
\end{equation}
wherein $\Lambda_w(\GNOG)$ is the weight lattice of the GNO-dual group $\GNOG$, 
and $\Wcal_{\GNOG}$ denotes the Weyl group. 
As shown in~\cite{Goddard:1976qe}, the lattice $\Lambda_w(\GNOG)$ coincides 
with the solutions of the generalised Dirac quantisation 
condition~\cite{Englert:1976ng}. 
The way $\HS_{\G}$ realises the Hilbert series is by counting BPS monopole 
operators, as studied by~\cite{tHooft:1977hy,Borokhov:2002cg,Borokhov:2002ib}. 
A key ingredient is the existence and uniqueness of the bare BPS 
monopole operator for each point in $\Lambda_w(\GNOG)$~\cite{Borokhov:2002cg}. 
These operators can be further characterised by their \emph{conformal 
dimension} $\Delta(m)$, which 
is given by~\cite{Borokhov:2002cg,Gaiotto:2008ak,Benna:2009xd,Bashkirov:2010kz}
\begin{equation}
 \Delta(m) = \frac{1}{2} \sum_{i=1}^n \sum_{\rho \in \mathcal{R}_i} |\rho(m)| - 
\sum_{\alpha\in \Phi_+} |\alpha(m)| \; . 
\label{eqn:def_conformal_dimension}
\end{equation}
Here, $\Phi_+$ denotes the set of positive roots of $\gfrak = \mathrm{Lie}(G)$, 
and $\mathcal{R}_i$ is the set of all weights of the $\G$-representation the 
$i$-th flavour of the $\Ncal=4$ hyper-multiplets transform in. 
We restrict our attention to \emph{good} theories in the sense 
of~\cite{Gaiotto:2008ak}, i.e.\ all non-trivial BPS monopole operators satisfy 
$\Delta>\frac{1}{2}$.
Lastly, compatibility with $\Ncal=4$ supersymmetry allows for a non-vanishing 
vacuum expectations value of a complex linear combination of the adjoint values 
scalar fields in the $\Ncal=4$ vector multiplet. The precise condition is that 
the vacuum expectation values can be any polynomial on the Lie algebra of the 
residual gauge group $\Hh_m = \mathrm{Stab}_\G(m)$ which has to be invariant 
under $\Hh_m$. This gives rise to the \emph{dressing factors} $P_{\G}(t;m)$ 
which are understood as \Poincare\ series of the algebra of $\Hh_m$ invariant 
polynomials on $\mathrm{Lie}(\Hh_m)$, see for instance~\cite[Sec.\ 
2]{Hanany:2016ezz}.
\paragraph{The matter fan and Hilbert bases}
In~\cite{Hanany:2016ezz} we introduced geometric concepts that, on the one 
hand, 
may simplify or at least systematise the computations for the monopole formula. 
On the other hand, the presented approach might lead to a better understand of 
the Coulomb branch itself.

The first step is just group and representation theory. The dominant Weyl 
chamber $\sigma$ of the GNO-dual group $\GNOG$ is a rational polyhedral cone 
inside a Cartan subalgebra $\tfrak$ of $\gfrak$. For the magnetic weights $m\in 
\Lambda_w(\GNOG) \subset \tfrak$ holds, while weights $\rho$ of $\G$ lie in 
the lattice $\Lambda_w(\G) \subset \tfrak^*$. Then, $\rho(m)$ is the dual 
pairing between dual spaces.
Next, we interpret the contributions to the conformal dimension as closed 
half-spaces and hyper-planes 
\begin{equation}
 H_{\rho}^{\pm} = \left\{ m \in \tfrak \, | \; \pm \rho(m) \geq 0 \right\}
 \und
 H_{\rho} = \left\{ m \in \tfrak \, | \;  \rho(m) = 0 \right\} \; ,
\end{equation}
for any $\rho \in \tfrak^*$.
Define the set $\Gamma$ of \emph{relevant weights}, i.e.\ those weights $\rho$ 
of $\G$ appearing in $\Delta$ for which neither $\rho$ nor $-\rho$ lies 
in the rational cone spanned by the simple roots $\Phi_s$ of $\G$. In other 
words, weights $\rho$ for which the intersection of $H_\rho$ with $\sigma$ is 
not only a face 
of $\sigma$.

The absolute values in~\eqref{eqn:def_conformal_dimension} are all 
simultaneously resolved on the rational polyhedral cones
\begin{equation}
 C_{\epsilon_1 , \ldots, \epsilon_{|\Gamma|}} \coloneqq \bigcap_{\rho \in 
\Gamma} H_{\rho}^{\epsilon_{\rho}} \with \epsilon_\rho = \pm \; .
\end{equation}
The action of the Weyl group $\Wcal_{\G}$ allows us to restrict to the dominant 
Weyl chamber $\sigma \subset \tfrak$, which introduces a collection 
\begin{equation}
 \tau_{\epsilon_1 , \ldots, \epsilon_{|\Gamma|}} \coloneqq C_{\epsilon_1 , 
\ldots, \epsilon_{|\Gamma|}} \cap \sigma 
\label{eqn:cones_cap_Weyl}
\end{equation}
of rational polyhedral cones inside the $\sigma$. The collection 
$\{\tau_{\epsilon_1 , \ldots, \epsilon_{|\Gamma|}}\}$ of theses cones, which is 
a finite set, generates a fan $F \subset \tfrak$. Since the contributing 
weights are only the relevant weights, the structure of $F$ is entirely 
determined by the hyper-multiplet matter content. Therefore, one could call $F$ 
the \emph{matter fan}.

As monopole operators are characterised by solutions to the generalised 
Dirac equation, one has to intersect the cones with the lattice 
$\Lambda_w(\GNOG)$, which yields
\begin{equation}
 S_\tau \coloneqq \tau \cap \Lambda_w(\GNOG) \for \tau \in F \; ,
\end{equation}

a collection of additive monoids in $\Lambda_w(\GNOG) $.
Under very reasonable assumptions, see further 
Sec.~\ref{subsec:product_groups}, all monoids are positive and finitely 
generated. For such monoids the minimal set of irreducible generators 
is the \emph{Hilbert basis}, denoted by $\Hilb(S_\tau)$.
\paragraph{Outline}
It is the purpose of this article to explore the implication of our 
approach~\cite{Hanany:2016ezz} for the algebraic properties of the monopole 
formula and to provide a treatment for arbitrary gauge groups.

We start with a brief exposition of the geometric and algebraic view on monoids 
and their associated algebras in Sec.~\ref{sec:monoid_algebras}. Key concepts 
will be triangulations of cones, refinements of fans, and free resolutions of 
lattice ideals.
Thereafter, Sec.~\ref{sec:Casi_invariance} is devoted to the Casimir 
invariance and exploits the algebraic properties of invariant polynomial 
algebras on Lie algebras.
With the collected mathematical background at hand, we explain the application 
to the monopole formula in Sec.~\ref{sec:application_monopole}. The main 
results lie in three different explicit expressions of the Hilbert series for 
the Coulomb branch of a theory with gauge group $\G$ and matter fan $F$.
In Sec.~\ref{sec:pole_structure} we prove the relation between the order of the 
pole of $\HS_{\G}(t)$ at $t=1$, $t\to \infty$ with the rank of the gauge 
group and, consequently, the dimension of the moduli space. In addition, we 
provide upper  bounds on the pole order for other roots of unity.
The understanding of the Hilbert bases and the Casimir invariants allows us to 
identify a sufficient set of generators for the chiral ring $\C[\MCoulomb]$ in 
Sec.~\ref{sec:chiral_ring}.
After the rather abstract considerations of the first sections, we comment on 
the application of the matter fan and the Hilbert bases to actual computations 
in Sec.~\ref{sec:implementation}. Firstly, we provide a recipe on how to 
compute the 
monopole formula using computer algebra software. Secondly, we illustrate the 
procedure for three quiver gauge theories.
In the end, Sec.~\ref{sec:conclusion} concludes.
For convenience of the reader, App.~\ref{app:AlgGeo} provides a reminder of 
employed algebro-geometric notions.
%%%%%%%%%%%%%%%%%%%%%%%%%%%%%%%%%%%%%%%%%%%%%%%%%%%%%%%%%%%%%%%%%%%%%%%%%%%%%%%%
   \section{Monoids and associated algebras}
\label{sec:monoid_algebras}
\subsection{Geometric picture}
We start by exploring some known properties of cones and monoids, for details 
see for instance~\cite{Ziegler:1995,Bruns:Herzog,Koch:2003,Bruns:2009}. Let $M$ 
denote an $r$-dimensional real vector space and $\Lambda$ an $r$-dimensional 
lattice in $M$. 
\subsubsection{Cones}
A \emph{rational polyhedral cone} $C$ is defined as all 
non-negative linear combinations of a finite set of vectors 
$X\coloneqq\{v_1,\ldots,v_s\} \subset \Lambda$, i.e.
\begin{equation}
 C\equiv \cone(X) = \left\{ \sum_{i=1}^s a_i v_i \, \big| \, a_i \in \R_{\geq0} 
\right\}
\end{equation}
We recall the following properties of a cone $C$:
\begin{itemize}
 \item $C$ is \emph{simplicial} if it is generated by linearly independent 
vectors.
 \item $C$ is \emph{strongly convex} or \emph{positive} if $C \cap (-C)= \{0\}$.
\end{itemize}

Two concepts are important to us: faces $F<C$ of a cone $C$ and the 
\emph{relative interior} $\Relint(C)$. Denote by $\Fcal(C)$ the set of faces of 
$C$, which is a finite set containing the two improper faces: $C$ and the 
trivial cone $\{0\}$. $\Fcal(C)$ is partially ordered with respect to 
inclusion.
\begin{myLemma}[{\cite[Cor.~2.7.6, p.~24]{Koch:2003}}]
 Every rational polyhedral cone $C$ is the disjoint union 
 \begin{equation}
  C = \biguplus_{F \in \Fcal(C)} \Relint(F)
 \end{equation}
of the relative interiors of all its faces.
\end{myLemma}
Moreover, for the strongly convex polyhedral cone 
$C=\cone(v_1,\ldots,v_s)$ the set $X=\{v_1,\ldots,v_s\}$ is a minimal 
generating set of $C$ if and only if $X$ contains exactly one vector $v_E$ from 
each edge $E$ of $C$ (and no other vectors). 
Later we will use a \emph{triangulation of a cone} $C$ which is a family 
$\Delta$ of finitely many simplicial sub-cones such that
\begin{enumerate}[(i)]
 \item $C= \cup_{\delta \in \Delta} \delta$,
 \item the faces of each $\delta \in \Delta$ are themselves members of 
$\Delta$, and
 \item the intersection of each pair $\delta,\epsilon\in \Delta$ is a face of 
both $\delta$ and $\epsilon$.
\end{enumerate}
For triangulations the following two lemmata hold:
\begin{myLemma}[{\cite[Lem.~2.11.1, p.~36]{Koch:2003}}]
\label{lem:properties_simplicial}
 For a simplicial cone $\delta$, generated by linearly independent vectors 
$x_1,\ldots,x_n$, the faces of $\delta$ are the simplicial cones $\cone(E)$, 
$E$ running through the subsets of $Y\coloneqq\{x_1,\ldots,x_n\}$. There is a 
bijection between the set of faces $\Fcal(\delta)$ and the power set 
$\mathcal{P}(Y)$.
\end{myLemma}
\begin{myLemma}[{\cite[Lem.~2.11.2, p.~36]{Koch:2003}}]
\label{lem:properties_triangle}
 Let $\Delta$ be a triangulation of the cone $C$, then the following holds:
 \begin{enumerate}[(i)]
  \item for each $\delta \in \Delta$: $\delta = \biguplus_{\substack{\epsilon 
\in \Delta \\ \epsilon \leq \delta}}\Relint(\epsilon)$,
  \item $C$ is the disjoint union $C = \biguplus_{\delta \in \Delta} 
\Relint(\delta)$.
 \end{enumerate}
\end{myLemma}
\begin{myLemma}[{\cite[Lem.~2.11.5, p.~37]{Koch:2003}}]
\label{lem:prefered_triangle}
 For each finitely generated cone $C$, there exists a triangulation $\Delta$ 
such that each simplicial sub-cone $\delta \in \Delta$ is generated by some of 
the generators of $C$.
\end{myLemma}
\subsubsection{Monoids}
Recall that the intersection of a polyhedral cone $C\subset M$ with the lattice 
$\Lambda$ yields a monoid $S_C = C \cap \Lambda$. Many concepts applicable for 
cones descend to the monoid such as the following:
\begin{itemize}
  \item The rank of a monoid $S_C$ equals the dimension of the cone $C$, i.e.\ 
$\rank(S_C) = \dim_{\R} (C)$.
 \item Let $F <C$ be a face of $C$ then $F_S \coloneqq F \cap S$ is called a 
face of $S$. The set of all faces is denoted as $\Fcal(S)\coloneqq \{F_S \, | 
\, F \in \Fcal(C)\}$, which is again a finite set.
 \item The relative interior of a face $F_S$ of $S$ is defined as $F_S 
\coloneqq S \cap \Relint(F) = F_S \cap \Relint(F)$.
  \item The set of invertible elements on $S$, denote by $S_0 =\{\nu \in S \, 
| \, -\nu \in S\}$, is the largest group contained in $S$. A monoid is called 
positive if $S_0=\{0\}$. (For our purposes, $S$ is positive if it descends from 
 a positive cone $C$.)
  \item For a positive monoid $S$, an element $\nu \in S$ is irreducible (in 
$S$) if $\nu = \nu_1+\nu_2$ for $\nu_1,\nu_2\in S$ is only possible if 
$\nu_1=0$ or $\nu_2=0$.
\end{itemize}
\begin{myLemma}[{\cite[Prop.~7.15, p.~137]{Miller:2005}}]
 For a positive monoid $S$, the following holds:
 \begin{enumerate}[(i)]
  \item $S$ has only finitely many irreducible elements.
  \item $S$ has a unique minimal generating set, given by the irreducible 
elements. This set is called Hilbert basis, denoted by $\Hilb(S)$.
 \end{enumerate}
\end{myLemma}
Note that the concept of a Hilbert basis relies on the positivity of the monoid 
as otherwise the irreducibilty is not well-defined.
For our set-up, Gordan's Lemma~\cite[Thm.~7.16, p.~137]{Miller:2005} ensures 
that the monoid $S_C$ is 
finitely generated. In the proof thereof, one shows that $S_C$ is generated 
by the finite set
\begin{equation}
\begin{aligned}
 &\{v_1,\ldots,v_s\} \cap \para{\Lambda}{v_1,\ldots,v_s} \\ 
 \with
 &\para{\Lambda}{v_1,\ldots,v_s} \coloneqq \Lambda \cap \{a_1 v_1 + 
\cdots + a_s v_s \, | \, a_i \in [0,1) \} \; ,
\end{aligned}
\end{equation}
where $\{v_1,\ldots,v_s\}$ are the minimal generators of $C$. The notion 
$\para{\Lambda}{v_1,\ldots,v_s}$ has been initiated in~\cite{Sebo:1990} and 
stems from the underlying parallelepiped spanned by $\{v_1,\ldots,v_s\}$.
Identifying the irreducible elements in $\para{\Lambda}{v_1,\ldots,v_s}$ 
together with the cone generators yields the elements of the Hilbert basis. 

Moreover, for a simplicial cone we obtain two 
different characterisations of $S_C$: firstly, as Minkowski sum
\begin{subequations}
\begin{equation}
 S_C = S_\mathrm{free} + \mathrm{par}_{\Lambda}(v_1,\ldots,v_s) \; ,
\end{equation}
and, secondly, as disjoint union
\begin{equation}
 S_C = \biguplus_{x \in \para{\Lambda}{v_1,\ldots,v_s} } (x + 
S_\mathrm{free})
\end{equation}
\end{subequations}
where $S_\mathrm{free}\coloneqq \spanN(v_1,\ldots,v_s) \subset S_C$ is the 
sub-monoid (freely) generated by the cone generators $\{v_1,\ldots,v_s\}$.
\subsection{Algebraic picture}
Having introduced monoids allows to discuss their associated algebras. Let 
$\K$ be a field and $S$ a monoid in the lattice $\Lambda$. By $\K[S]$ we denote 
the monoid algebra which is a $\K$-vector space with basis $\bt^\nu$, for $\nu 
\in S$. A generic element of $\K[S]$ is of the form $a_1 \bt^{\nu_1}+\cdots + 
a_m \bt^{\nu_m}$ for $m\in \NN$, $a_i \in \K$, $\nu_i \in S$. The additive 
structure of $\K[S]$ is clear and multiplication arises via $\bt^{\nu_1} \cdot 
\bt^{\nu_2} = \bt^{\nu_1 + \nu_2}$.
\begin{myLemma}[{\cite[Prop.~2.7, p.~54 \& Prop.~4.22, p.~137]{Bruns:2009}}]
 For a monoid $S$ of the lattice $\Lambda$, and $\K[S]$ the monoid ring 
(with $\K$ a field).  The (Krull) dimension is given by $\dim (\K[S])= 
\rank(S)$. 
In addition, the following are equivalent:
\begin{enumerate}[(i)]
 \item $S$ is a finitely generated monoid.
 \item $\K[S]$ is a finitely generated $\K$-algebra.
\end{enumerate}%
\end{myLemma}%
Let us briefly recall gradings of monoids, rings, and modules. Let $A$ be a 
monoid, then:
\begin{itemize}
 \item An $A$-graded ring is a ring $R$, together with the decomposition $R= 
\bigoplus_{a \in A} R_a$ such that $R_a \cdot R_{b} \subseteq R_{a+b}$, for all 
$a,b\in A$. The grading is positive if the only elements of $R$ with degree 
$0\in A$ are constants, i.e.\ $R_0 = \K$ and $\K$ a field.
\item An $A$-graded $R$-module is an $R$-module $M$, together with a 
decomposition $M= \bigoplus_{a\in A}M_a$ such that $R_a \cdot M_b \subseteq 
M_{a+b}$ for all $a,b\in A$
  \item An $A$-graded monoid is a monoid $S$, together with a disjoint 
decomposition $S=\biguplus_{a\in A} S_a$ such that $S_a + S_{b} \subseteq 
S_{a+b}$ for all $a,b \in A$. We introduce the following monoid homomorphism
\begin{equation}
 \phi : \begin{matrix} S& \to& A  \\
         \nu &\mapsto& \phi(\nu)
        \end{matrix}
        \st S_a = \{\nu\in S \, | \,\phi(\nu)=a   \} \; ,
\end{equation}
which satisfies $\phi(\nu_1 +\nu_2) = \phi(\nu_1) + \phi(\nu_2)$ and 
$\phi(n\cdot \nu) = n \cdot \phi(\nu)$ for any $n\in \NN$ and 
$\nu, \nu_1,\nu_2\in S$.
The grading is positive if $S_0 =\{0\}$.
\end{itemize}
For a monoid $S$, a subset $T\subset S$ is an ideal (of $S$) if $S+I \subseteq 
I$. The radical of an ideal $I$ is $\mathrm{Rad}(I)\coloneqq \{c \, | \, m 
\cdot c \in I \text{ for some } m\in \NN\}$, and $I$ is a radical ideal if $I = 
\mathrm{Rad}(I)$.
Also, a subset $T \subseteq \Lambda$ is called an $S$-module if $S+T 
\subseteq T$. We have the following statements:
\begin{myLemma}[{\cite[Prop.~6.1.1, p.~257]{Bruns:Herzog}}]
\label{lem:relations_ideals_modules}
 Let $I$ be an arbitrary subset of $S$: $I$ is an ideal in $S$ if and only if 
$\K[I]$ is an ideal in $\K[S]$. A subset $T\subseteq \Lambda$ is an $S$-module 
if and only if the $\K$-vector space $\K(T)$, generated by $\bt^\nu$, $\nu\in 
T$, in $\K[\Lambda]$ is a $\K[S]$-module.
\end{myLemma}
This correspondence between ideal and module in $S$ and $\K[S]$ extends to the 
notion of radical ideals, prim ideals etc. Important results are  (see 
App.~\ref{app:AlgGeo} for normality and Cohen-Macaulay)
\begin{myLemma}[{\cite[Lem.~6.1.6, p.~261]{Bruns:Herzog}}]
 Let $S$ be a monoid, then the ideal generated by the elements $\bt^\nu$, $\nu 
\in\Relint(S)$ is a radical ideal, and is contained in every non-zero graded 
radical ideal of $\K[S]$.
\end{myLemma}
\begin{myLemma}[{\cite[Cor.~2.24, p.~61]{Bruns:2009}}]
 $S_C=C\cap \Lambda$ is always a normal monoid, provided $C$ is finitely 
generated.
\end{myLemma}
\begin{myLemma}[{\cite[Thm.~6.1.4, p.~260]{Bruns:Herzog}}]
\label{lem:normal_algebra}
 For $S$ a monoid and $\K$ a field, the following are equivalent:
 \begin{enumerate}[(i)]
  \item $S$ is a normal monoid.
  \item $\K[S]$ is normal.
 \end{enumerate}
\end{myLemma}
\begin{myLemma}[{\cite[Thm.~6.3.5, p.~272]{Bruns:Herzog}}]
\label{lem:canonical_module}
 Let $S$ be a normal monoid and $\K$ a field, then
 \begin{enumerate}[(i)]
  \item $\K[S]$ is a Cohen-Macaulay ring and
  \item the ideal $I$ generated by the monomials $\bt^\nu$ with $\nu \in 
\Relint(S)$ is the canonical module of $\K[S]$.
 \end{enumerate}
\end{myLemma}
\subsection{Hilbert series}
As usual, one defines the Hilbert series for a positively $A$-graded $R$-module 
$M$ via
\begin{equation}
 \HilbS_M(\bt) \coloneqq \sum_{a \in A} \HilbS(M,a)\cdot \bt^a \with 
\HilbS(M,a) \coloneqq \dim_{\K}M_a \; . \label{eqn:def_Hilbert_series}
\end{equation}
The Hilbert series for a positively $A$-graded $S$-module $T$ is defined 
similarly 
\begin{subequations}
\begin{equation}
\label{eqn:HilbS_module_monoid_1}
 \HilbS_T(\bt) \coloneqq \sum_{a \in A}\HilbS(T,a)\cdot \bt^a \with 
\HilbS(S,a) \coloneqq \mathrm{card}(T_a) \; ,
\end{equation}
where $\mathrm{card}(T_a)$ is defined as the number of points in $T_a$. 
Note that positivity of the $A$-grading implies that both $\HilbS(M,a)$ and 
$\HilbS(S,a)$ are finite for any $a\in A$.
Moreover, if follows from Lem.~\ref{lem:relations_ideals_modules} that 
$\HilbS_{T}(\bt)$ equals $\HilbS_{\K(T)}(\bt)$, since $\K(T)$ is a 
$\K[S]$-module.
There exists an alternative characterisations 
of~\eqref{eqn:HilbS_module_monoid_1} which counts each point in the $S$-module 
$T$ precisely once according to their $A$-grading
\begin{equation}
 \HilbS_T(\bt) = \sum_{\nu \in T}\bt^{\phi(\nu)} \; .
\end{equation}
\label{eqn:HilbS_module_monoid}
\end{subequations}
This is the very same spirit as the monopole formula 
of~\cite{Cremonesi:2013lqa} which counts each bare monopole operator instead of 
providing the Hilbert function $\HilbS(\G,d)\equiv \{m \, | \, \Delta(m)=d\}$ 
for the Coulomb branch.

Suppose $S_C=C\cap \Lambda$ is the associated monoid for the cone 
$C=\cone(X)$, with $X=\{v_1,\ldots,v_s\}$ the cone generators. We would like to 
compute two objects, the Hilbert 
series of $S_C$ and $\Relint(S_C)$. For this, we 
follow~\cite{Bruns:Herzog,Koch:2003,Bruns:2009}.
\subsubsection{Hilbert series for a monoid}
Let $\Delta$ be a triangulation of $C$, then define $\omega_\delta \coloneqq 
\Lambda \cap \Relint( \delta)$ for each $\delta \in \Delta$. The immediate 
consequence of Lem.~\ref{lem:properties_triangle} for the monoid $S=\Lambda 
\cap C$ is
\begin{equation}
 \HilbS_{S}(\bt) = \sum_{\delta \in \Delta} \HilbS_{\omega_\delta} (\bt) \;.
\end{equation}
Since every $\delta$ is simplicial and we can choose $\Delta$ as in 
Lem.~\ref{lem:prefered_triangle}. The set of minimal generators of $\delta$ is 
denoted 
by $X_\delta \equiv \{y_1,\ldots,y_r\} \subset X$, a subset of 
the cone generators $X$ of $C$. Then 
$X_\delta$ generates a free monoid $\sigma_\delta = \spanN(X_\delta)$ whose 
Hilbert series reads
\begin{equation}
 \HilbS_{\sigma_\delta} (\bt)= \frac{1}{\prod_{y \in X_\delta} 
\left(1-\bt^{\phi(y)} \right)} \; .
\end{equation}
For $\omega_\delta$ we obtain the disjoint decomposition
\begin{equation}
\begin{aligned}
 \omega_\delta &= \biguplus_{x \in \paraprim{\Lambda}{X_\delta}} (x+ 
\sigma_\delta) \\
\with \paraprim{\Lambda}{X_\delta} &\coloneqq \Lambda \cap \left\{ q_1 y_1 + 
\cdots q_r y_r \, | \, q_i \in (0,1]  \right\}
\end{aligned}
\end{equation}
Thus, we arrive at
\begin{equation}
 \HilbS_{\omega_\delta}(\bt) = \frac{\sum_{x\in 
\paraprim{\Lambda}{X_\delta}}  \bt^{\phi(x)}}{\prod_{y \in X_\delta} 
\left(1-\bt^{\phi(y)} \right)} \; . \label{eqn:HS_Relint_simplicial_cones}
\end{equation}
Alternative, we rewrite $\paraprim{\Lambda}{X_\delta}$ in slices $B_a$ of 
constant degree $a$, i.e.
\begin{equation}
 \paraprim{\Lambda}{X_\delta}  \equiv \bigoplus_{a\in A'_\delta} B_a \; ,
\end{equation}
labelled by a finite subset $A'_\delta\subset A$. 
Rewriting~\eqref{eqn:HS_Relint_simplicial_cones} yields 
\begin{equation}
 \HilbS_{\omega_\delta}(\bt) = \frac{\sum_{a \in A'_\delta}  
\mathrm{card}(B_a)  \bt^{a}}{\prod_{y \in X_\delta} 
\left(1-\bt^{\phi(y)} \right)} \; .
\end{equation}
Finally, we can summarise
\begin{equation}
 \HilbS_{S}(\bt) = \sum_{\delta \in \Delta} \frac{\sum_{x\in 
\paraprim{\Lambda}{X_\delta}}  \bt^{\phi(x)}}{\prod_{y \in X_\delta} 
\left(1-\bt^{\phi(y)} \right)} 
= \sum_{\delta \in \Delta} \frac{\sum_{a \in A'_\delta}  
\mathrm{card}(B_a)  \bt^{a}}{\prod_{y \in X_\delta} 
\left(1-\bt^{\phi(y)} \right)} \; .
\label{eqn:HilbSeries_monoid}
\end{equation}
The two representations~\eqref{eqn:HilbSeries_monoid} are based on the two 
different counting arrangements of~\eqref{eqn:HilbS_module_monoid}.
\subsubsection{Hilbert series for the relative interior of a monoid}
Computing the Hilbert series for the radical ideal $\Relint(S_C)$ proceeds 
essentially similar, but we need to modify the triangulation. Choosing the 
triangulation $\Delta$ of $C$ as in Lem.~\ref{lem:prefered_triangle}, the 
triangulation $\Delta$ induces also triangulations for all faces of $C$. Thus, 
we can define a subset 
\begin{equation}
 \Delta'\coloneqq \Delta \setminus \left\{ \tau \in \Delta \, | \, \exists 
\sigma \in \Fcal(C)\setminus C \st \tau \subset \sigma \right\}  \subset \Delta 
\; ,
\end{equation}
which then yields the desired property
\begin{equation}
 \Relint(C) = \biguplus_{\delta \in \Delta'} \Relint(\delta) \; .
\end{equation}
Consequently, we obtain the Hilbert series by restriction of the 
results~\eqref{eqn:HilbSeries_monoid} as
\begin{equation}
 \HilbS_{\Relint(S)}(\bt) = \sum_{\delta \in \Delta'} \frac{\sum_{x\in 
\paraprim{\Lambda}{X_\delta}}  \bt^{\phi(x)}}{\prod_{x \in X_\delta} 
\left(1-\bt^{\phi(x)} \right)} 
= \sum_{\delta \in \Delta'} \frac{\sum_{a \in A'_\delta}  
\mathrm{card}(B_a)  \bt^{a}}{\prod_{x \in X_\delta} 
\left(1-\bt^{\phi(x)} \right)} \; .
\label{eqn:HilbSeries_relint}
\end{equation}
\subsection{Free resolution}
\label{subsec:resolutions}
Besides the pure computation of Hilbert series for monoid rings, we can 
additionally shed light on algebraic properties. We follow~\cite{Miller:2005}. 
Suppose $S$ is a monoid in $\Lambda$ and $\phi: S \to A$ is the monoid morphism 
providing the $A$-grading of $S$. For our intends and purposes, $S$ is the 
monoid associated to a positive rational polyhedral cone and, hence, by 
Gordan's lemma is finitely generated by $\Hilb(S)= \{y_1,\ldots,y_n\}$. 
It follows that $\phi(\Hilb(S))$ generates the semi-group $\phi(A)\subset A$, 
but is not necessarily a minimal set. Consider as example $A=\NN$ and $\phi: S 
\to \NN$, but $\phi(S)\subset \NN$ can be generated by any number $q$ of 
elements, for $1\leq q \leq n$ depending on the degrees of $\phi(y_i)$.
Another example is the grading of $S$ by itself, i.e. $A=S$ and $\phi= 
\mathrm{id}|_{S}$. Then $\Hilb(S)$, of course, remains the minimal generating 
set.

Let $e_i$ for $i=1,\ldots,n$ be the standard basis of $\Z^n$ and define the 
group homomorphism
\begin{equation}
 \Phi: \begin{matrix} \Z^n & \to & \phi(S) \\ e_i & \mapsto  & \phi(y_i)        
       \end{matrix} \; .
\end{equation}
The kernel $L\coloneqq \Kern(\Phi)$ of this map is a lattice in $\Z^n$. 
Consider the $A$-graded polynomial ring $R=\K[z_1,\ldots,z_n]$, where the 
grading is given by a degree map $\deg:\Z^n \to A$ such that a monomial 
$\bz^u = z_1^{u_1} \cdots z_n^{u_n}$, for $u\in \NN^n$, has degree 
$\deg(\bz^u)=\deg(u)= \ba \in A$. The distinguished set of $n$ 
elements $\deg(z_1),\ldots,\deg(z_n)$ in $A$ is denoted by $\ba_1,\ldots,\ba_n$.

Define the so-called \emph{lattice ideal} $I_L \subset R$ associated to $L$ via
\begin{equation}
 I_L = \langle \bz^u - \bz^v \, | \, u,v \in \NN^n \with u -v \in L 
\rangle \; .
\end{equation}
\begin{myLemma}[{\cite[Thm.~7.3, p.~130]{Miller:2005}}]
 The $A$-graded monoid ring $\K[S]$ is isomorphic to the quotient $R \slash 
I_L$.
\end{myLemma}
Employing the additivity of the Hilbert series in exact sequences, 
i.e.
\begin{equation}
 0\to A \to B \to C \to 0 \, , \text{ then } \quad \HilbS_B = \HilbS_A + 
\HilbS_C \; ,
\end{equation}
we can deduce the Hilbert series of the monoid algebra via
\begin{equation}
 0 \to I_L \to R \to R \slash I_L \to 0 \, , \text{ hence } \HilbS_{\K[S]} = 
\HilbS_{R} - \HilbS_{I_L} \; .
\end{equation}
Since the Hilbert series of the $A$-graded polynomial ring $R$ is simply given 
by
\begin{equation}
 \HilbS_R(\bt) = \frac{1}{\prod_{i=1}^n (1-\bt^{\ba_i})} \; ,
\end{equation}
the question arises for the Hilbert series of the lattice ideal $I_L$. 
Here, we can employ the notion of a \emph{free resolution} of an $R$-module 
$M$. 
\subsubsection{Free resolution}
For the moment, we do not consider graded rings or modules. Recall that a free 
$R$-module $F$ is a direct sum $F\cong R^r$ for some $r\in \NN$. A free 
resolution of a finitely generated $R$-module $M$ is an exact sequence of the 
form
\begin{equation}
  \cdots \xrightarrow{} F_i \xrightarrow{f_i} F_{i-1} \xrightarrow{} \cdots 
\xrightarrow{f_3} F_2 \xrightarrow{f_2} F_1 \xrightarrow{f_1} F_0 
\xrightarrow{f_0} M \xrightarrow{} 0  
\end{equation}
 between finitely generated free $R$-modules $F_i$. The resolution is finite if 
there exists some $l\in \NN$ such that $F_l\neq 0$ and $F_j=0$ for $j>l$. If 
that is the case, the resolution is of length $l$ and is of the form
\begin{equation}
 0 \rightarrow{} F_l \xrightarrow{f_l} F_{l-1} \xrightarrow{} \cdots 
\xrightarrow{f_3} F_2 \xrightarrow{f_2} F_1 \xrightarrow{f_1} F_0 
\xrightarrow{f_0} M \xrightarrow{} 0  \; .
\end{equation}
\begin{myLemma}[Hilbert Syzygy Theorem, {\cite[Thm.~2.1, p.~259]{Cox:Little}}]
 Let $R=\K[y_1,\ldots,y_n]$. Then every finitely generated $R$-module has a 
finite free resolution of length at most $n$.
\end{myLemma}
Benefit of a free resolution is that all appearing $R$-modules $F_i$ are free 
and, hence, the Hilbert series of each $F_i$ is simple to compute.
Quite important, the free resolution keeps track of relations between 
generators of modules. To be more precise, let $M \subset R^m$ be an $R$-module 
generated by the set $F=\{f_1,\ldots,f_s\}$. The syzygy module 
$\mathrm{Syz}(F)$ of $F$ is the set
\begin{equation}
 \mathrm{Syz}(F)= \left\{ (g_1,\ldots,g_s) \in R^s \, | \, f_1 g_1 + \cdots f_s 
g_s =0  \right\} \; .
\end{equation}
An element of $\mathrm{Syz}(F)$ is called a syzygy. Put differently, the syzygy 
module of $F$ is the kernel of the map $R^s \to M$ determined by $e_i \mapsto 
f_i$, where $e_i$ is the standard basis of $R^s$.
\subsubsection{Graded free resolution}
For a positively $A$-graded $R$-module we need to introduce the notion of 
graded homomorphisms. Let $M$, $N$ be graded modules over 
$R=\K[z_1,\ldots,z_n]$. A homomorphism $f: M \to N$ is a graded homomorphism of 
degree $\bg \in A$ if $f(M_{\ba}) \subset N_{\ba+\bg}$ for all $\ba\in A$.
Moreover, we define the translate $M(\bg)$ of the $R$-module $M$ by $\bg \in A$ 
via the direct sum
\begin{equation}
 M(\bg)= \bigoplus_{\ba \in A} M(\bg)_{\ba} \with M(\bg)_{\ba} \coloneqq 
M_{\bg+\ba} \; .
\end{equation}
Then $M(d)$ is a graded $R$-module. Now, for a positively $A$-graded $R$-module 
$M$ a graded resolution of $M$ is a resolution of the form 
\begin{equation}
 \cdots \xrightarrow{} F_2 \xrightarrow{f_2} F_1 \xrightarrow{f_1} F_0 
\xrightarrow{f_0} M \xrightarrow{} 0  \; ,
\label{eqn:graded_resolution}
\end{equation}
where the $F_i$ are (twisted) free graded modules of the form 
$R(-\bg_{i,1})\oplus \cdots \oplus R(-\bg_{i,p})$, and the homomorphisms 
between them are graded of degree $0$.
\begin{myLemma}[graded Hilbert Syzygy Theorem, {\cite[Thm.~3.8, 
p.~271]{Cox:Little}}]
 Let $R=\K[z_1,\ldots,z_n]$. Then every finitely generated graded 
$R$-module has a finite graded resolution of length at most $n$.
\end{myLemma}
For a free $R$-module $F=R(-\bg_{1})\oplus \cdots \oplus R(-\bg_{p})$ the 
Hilbert series is simple to evaluate
\begin{align}
 \HilbS_F(\bt) = \sum_{j=1}^p \HilbS_{R(-\bg_j)}(\bt) = \sum_{j=1}^p 
\bt^{\bg_j} \cdot
\HilbS_{R} (\bt)= \frac{\sum_{j=1}^p \bt^{\bg_j}}{\prod_{i=1}^n 
(1-\bt^{\ba_i})} \; .
\end{align}
Consequently, the Hilbert series for an $R$-module $M$ with free 
resolution~\eqref{eqn:graded_resolution} of length $l$ computes to
\begin{equation}
 \HilbS_{M}(\bt) = \sum_{i=0}^l (-1)^i \HilbS_{F_i}(\bt) = 
 \frac{\sum_{i=0}^l (-1)^i \sum_{j=1}^{p_i} \bt^{\bg_{i,j}} }{\prod_{i=1}^n ( 
1-\bt^{\ba_i}) } \equiv \frac{\Kcal(M,\bt)}{\prod_{i=1}^n ( 
1-\bt^{\ba_i})}
\end{equation}
wherein the $\Kcal(M,\bt)$ denotes the so-called K-polynomial.

Ultimately, we desire to compute the Hilbert series of the quotient 
$\K[S]\cong R\slash I_L$, all we have to do is to evaluate the free resolution 
of $I_L$. The result is then given by
\begin{equation}
 \HilbS_{\K[S]}(\bt) = \frac{1-\Kcal(I_L,\bt)}{\prod_{i=1}^n ( 1-\bt^{\ba_i})} 
\; .
\end{equation}
We will illustrate the practical application of the result later.
% 
%%%%%%%%%%%%%%%%%%%%%%%%%%%%%%%%%%%%%%%%%%%%%%%%%%%%%%%%%%%%%%%% 
%%%%%%%%%%%%%%%%%%%%%%%%%%%%%%%%%%%%%%%%%%%%%%%%%%%%%%%%%%%%%%%%
%
\subsection{Canonical module}
As introduced in Lem.~\ref{lem:canonical_module}, the relative interior of a 
normal monoid gives rise to the canonical module of a Cohen-Macaulay ring. The 
canonical module captures important properties of Cohen-Macaulay rings and is 
intimately related to Gorenstein rings. We refer to~\cite{Bruns:2009} 
for the details and provide the relevant points here. 

First of all, a canonical module is by definition a Cohen-Macaulay module, i.e. 
satisfies the equality between depth and dimension, c.f.\ 
App.~\ref{app:AlgGeo}. Moreover, a result by 
Stanley relates the Hilbert series between canonical module and 
Cohen-Macaulay ring:
\begin{myLemma}[{\cite[Thm.~6.40, p.~232]{Bruns:2009}}]
\label{lem:Stanley}
Let $S$ be a positive affine monoid and $R$ a finitely generated 
Cohen-Macaulay $S$-graded $\K$-algebra, with canonical module $\Omega_R$. Then
\begin{equation}
 \HilbS_{\Omega_R}(\bt) = (-1)^d \HilbS_{R}(\bt^{-1}) \; , \qquad d = \dim(R)\; 
.
\end{equation}
\end{myLemma}
A Cohen-Macaulay ring $R$ is Gorenstein if it is isomorphic to its 
own canonical module, i.e.\ $\Omega_R \cong R(a)$ for some $a\in S$. 
Consequently, the Hilbert series satisfies
\begin{equation}
 \HilbS_{R}(\bt^{-1}) \stackrel{\text{Lem.}~\ref{lem:Stanley} }{=}(-1)^d  
\HilbS_{\Omega_R}(\bt) = (-1)^{d} \bt^{-a} \HilbS_R(\bt) \; , 
\end{equation}
which is precisely the known criterion by Stanley~\cite{Stanley:1978} for $R$ 
to be Gorenstein.
Specialising to the case of normal monoids, one obtains
\begin{myLemma}[{\cite[Thm.~6.49, p.~237]{Bruns:2009}}]
 Let $S$ be a normal affine monoid of rank $d$, and let $x_1,\ldots, x_k$ be 
the cone generators of $C_S=\mathrm{Span}_{\R_{\geq0}}(S)$. Then
\begin{subequations}
\begin{equation}
 \HilbS_{\K[S]}(\bt) = \frac{Q(\bt)}{\prod_{i=1}^k (1-\bt^{x_i})}
 \label{eqn:HS_monoid}
\end{equation}
with a integer valued polynomial $Q(t)$. The Hilbert series for the ideal 
associated to $\Relint(S)$ is given by
\begin{equation}
 \HilbS_{\K[\Relint(S)]}(\bt) = (-1)^{k-d}\; \frac{\bt^{\sum_{j=1}^k x_j} \, 
Q(\bt^{-1})}{\prod_{i=1}^k (1-\bt^{x_i})} \; .
\end{equation}
\end{subequations}
\end{myLemma}
The relation to Gorenstein rings is easily established for monoid 
algebras~\cite[Thm.~6.32, p.~226]{Bruns:2009}, 
because for a normal monoid $S$, $\K[S]$ is Gorenstein if and only if there 
exists $x \in S$ such that $\Relint(S)= x + S$. Explicitly, one observe 
$\K[\Relint(S)] \cong \K[S](x)$, such that the Hilbert series for the ideal has 
a palindromic numerator, since the monoid ring has. 
% %%%%%%%%%%%%%%%%%%%%%%%%%%%%%%%%%%%%%%%%%%%%%%%%%%%%%%%%%%%%%%%%%%%%%%%%%%%%%%%%
   \section{Casimir invariance}
\label{sec:Casi_invariance}
Previously~\cite{Hanany:2016ezz}, we have seen that the number of dressed 
monopole operators for a 
magnetic charge $m$ is determined by the ration 
$\frac{P_{\G}(t;m)}{P_{\G}(t;0)}$. Let us study this ratio in more 
detail. To start with, the classical dressing factors $P_{\G}(t;m)$ had been 
identified with the \Poincare\ series $\PoinS(t)$ of invariant polynomial 
algebras:
\begin{equation}
 P_{\G}(t;m) = \PoinS_{\mathfrak{J}(\hfrak_m)^{\Wcal_{\Hh_m}}}(t)= 
\frac{1}{\prod_j(1-t^{b_j})} 
\quad \text{and} \quad 
 P_{\G}(t;0) = \PoinS_{\mathfrak{J}(\hfrak)^{\Wcal_{\G}}} (t)= 
\frac{1}{\prod_j(1-t^{d_j})} \; ,
\label{eqn:def_Dress_factors}
\end{equation}
wherein $\hfrak_m$ is the Lie algebra of the stabiliser $\Hh_m$ of $m$ in $\G$, 
and $\Wcal_{\Hh_m}$ is the Weyl group of $\Hh_m$. The ring 
$\mathfrak{J}(\hfrak_m)^{\Wcal_{\Hh_m}}$ of invariants is Gorenstein for any 
$m$.
Moreover, we know $\T\subseteq \Hh_m \subseteq \G$ and 
$\{1\}\subseteq \Wcal_{\Hh_m} 
\subseteq \Wcal_{\G}$, with $\T$ a maximal torus of $\G$ and $\tfrak = 
\mathrm{Lie}(\T)$. Without loss of generality we can choose all Cartan 
sub-algebras $\hfrak_m$ and $\hfrak$ to be equal to $\tfrak$; thus, we deduce
\begin{equation}
 \mathfrak{J}(\hfrak)^{\Wcal_{\G}} \subset 
\mathfrak{J}(\hfrak_m)^{\Wcal_{\Hh_m}} \subset \mathfrak{J}(\tfrak)^{\{1\}} 
\equiv \mathfrak{P}(\tfrak) \; .
\end{equation}
Here, $\mathfrak{P}(\tfrak)$ denotes the algebra of polynomials on $\tfrak$. 
We also know that all algebras are finitely generated by $r=\rank(\G)$ 
algebraically independent, homogeneous elements, i.e.
\begin{subequations}
 \begin{alignat}{3}
 \mathfrak{J}(\hfrak)^{\Wcal_{\G}} &\cong \C[f_1,\ldots,f_r] & 
 \quad  &\text{with} \quad & 
\deg(f_j)&=d_j \; , \\
\mathfrak{J}(\hfrak_m)^{\Wcal_{\Hh_m}} &\cong \C[h_1,\ldots,h_r] &
\quad &\text{with} \quad  &
\deg(h_j)&=b_j \; .
\end{alignat}
\end{subequations}
Now, consider the ideal 
\begin{equation}
I_{\G}^{\mathrm{max}}\equiv \langle f_1,\ldots, f_R \rangle \coloneqq
\left\{ \sum 
\chi_j f_j \, | \, \chi_j \in \mathfrak{J}(\hfrak)^{\Wcal_{\G}} \right\}
\subset \mathfrak{J}(\hfrak)^{\Wcal_{\G}} \; ,
\end{equation}
which is a maximal ideal of $\mathfrak{J}(\hfrak)^{\Wcal_{\G}}$ by results of 
Hilbert. However, $I_{\G}^{\mathrm{max}}$ is not an ideal in 
$\mathfrak{J}(\hfrak_m)^{\Wcal_{\Hh_m}}$, but it can be promoted to one, 
by considering the ideal 
\begin{equation}
 \widetilde{I}_{\G}^{\mathrm{max}} \equiv \langle f_1,\ldots, f_R \rangle 
\coloneqq \left\{ \sum 
\phi_j f_j \, | \, \phi_j \in \mathfrak{J}(\hfrak_m)^{\Wcal_{\Hh_m}} \right\}
\subset \mathfrak{J}(\hfrak_m)^{\Wcal_{\Hh_m}} \; ,
\end{equation}
which is spanned by the same elements.
Then we claim
\begin{myProp}
\label{prop:Casimir_inv}
 \begin{enumerate}[(i)]
 \item The module $M_m^{\mathrm{Dress}}\coloneqq 
\mathfrak{J}(\hfrak_m)^{\Wcal_{\Hh_m}} \slash \widetilde{I}_{\G}^{\mathrm{max}}$ 
is 
precisely what is expected from the physical picture of the dressed monopole 
operators of charge $m$. That is, the bare monopole $V_m$ can be dressed by 
polynomials in the Casimir invariants of $\Hh_m$ which do not contain the 
Casimir invariant of the entire gauge group $\G$.
  \item The ratio $\frac{P_{\G}(t;m)}{P_{\G}(t;0)}$ of dressing factors equals 
the \Poincare\ series of the quotient $M_m^{\mathrm{Dress}}$.
  \item By our previous results~\cite[App. A]{Hanany:2016ezz}, we know 
  \begin{equation}
\lim_{t\to 1} \PoinS_{M_m^{\mathrm{Dress}}}(t) = 
\frac{\left|\Wcal_{\G} \right|}{\left|\Wcal_{\Hh_m} \right|} \in \mathbb{N} \; .
\label{eqn:HS_limit}
  \end{equation}
By definition of a \Poincare\ series, which is morally the same as a Hilbert 
series~\eqref{eqn:def_Hilbert_series}, the 
fact~\eqref{eqn:HS_limit} 
implies that $M_m^{\mathrm{Dress}}$ is a 
finitely graded and finite dimensional module; i.e. 
$\dim(M_m^{\mathrm{Dress}})=\lim_{t\to1}\PoinS_{M_m^{\mathrm{Dress}}}(t)=\sum_{
i\in \mathbb{N}} \dim((M_m^{\mathrm{Dress}})_i)<\infty$. Herein, 
$(M_m^{\mathrm{Dress}})_i $ equals the number of dressed monopole operators of 
charge $m$ with degree $\Delta(m)+i$.
  \item The construction of the module $M_m^{\mathrm{Dress}}$ as quotient 
together with the finite limit~\eqref{eqn:HS_limit} and the definition of the 
\Poincare\ series proves that $\frac{P_{\G}(t;m)}{P_{\G}(t;0)}$ is indeed an 
ordinary polynomial in $t$ for any $m$.
  \item $\PoinS_{M_m^{\mathrm{Dress}}}(t)$ is a palindromic polynomial for any 
$m$. The order of $\PoinS_{M_m^{\mathrm{Dress}}}(t)$ equals the difference in 
number of reflections in $\Wcal_\G$ and number of reflections on 
$\Wcal_{\Hh_m}$.
\end{enumerate}
\end{myProp}
We start with the \Poincare\ series of the quotient 
$\mathfrak{J}(\hfrak_m)^{\Wcal_{\Hh_m}} \slash 
\widetilde{I}_{\G}^{\mathrm{max}}$.
Starting by induction, for $\widetilde{I}_j \equiv \langle f_j \rangle 
\subset \C[h_1,\ldots,h_R]$ we obtain the exact sequence of 
$\mathfrak{J}(\hfrak_m)^{\Wcal_{\Hh_m}}$-modules
\begin{equation}
 0 \rightarrow \langle f_j \rangle \rightarrow \C[h_1,\ldots,h_R] \rightarrow 
 \C[h_1,\ldots,h_R] \slash \langle f_j \rangle \rightarrow 0
\end{equation}
such that the \Poincare\ series behaves as 
\begin{subequations}
\begin{align}
 \PoinS_{\C[h_1,\ldots,h_R]}(t) &= \PoinS_{\langle f_j \rangle}(t)  
+\PoinS_{\C[h_1,\ldots,h_R] \slash \langle f_j \rangle}(t) \\
\text{with } \qquad &\PoinS_{\C[h_1,\ldots,h_R]}(t) = \frac{1}{\prod_j 
(1-t^{b_j})} \\
\text{and } \qquad &\PoinS_{\langle f_j \rangle}(t) = t^{d_j}\cdot 
\PoinS_{\C[h_1,\ldots,h_R]}(t) \\
\Longrightarrow \qquad  &\PoinS_{\C[h_1,\ldots,h_R] \slash \langle f_j \rangle} 
(t) = (1-t^{d_j}) \cdot \PoinS_{\C[h_1,\ldots,h_R]}(t)
\end{align}
\end{subequations}
Since the generators $\{f_j\}$ are algebraically independent, we can proceed 
iteratively and obtain the exact sequence of 
$\mathfrak{J}(\hfrak_m)^{\Wcal_{\Hh_m}}$-modules
\begin{equation}
 0 \rightarrow \langle f_1,\ldots,f_R \rangle \rightarrow \C[h_1,\ldots,h_R] 
\rightarrow 
 \C[h_1,\ldots,h_R] \slash \langle f_1,\ldots,f_R \rangle \rightarrow 0 \; .
\end{equation}
The \Poincare\ series then reduces to
\begin{equation}
 \PoinS_{ \C[h_1,\ldots,h_R] \slash \langle f_1,\ldots,f_R \rangle}(t) = 
\prod_{j=1}^R (1-t^{d_j}) \cdot \PoinS_{\C[h_1,\ldots,h_R]}(t) \; . 
\label{eqn:HS_quotient}
\end{equation}
Employing~\eqref{eqn:def_Dress_factors} in~\eqref{eqn:HS_quotient} yields
\begin{equation}
 \PoinS_{ M_m^{\mathrm{Dress}} }(t) = \frac{P_{\G}(t;m)}{P_{\G}(t;0)} 
\end{equation}
and taking the limit $t\to 1$ yields~\eqref{eqn:HS_limit}. Next, the 
palindromic character of $\PoinS_{ M_m^{\mathrm{Dress}} }(t)$ is proven as 
follows:
\begin{equation}
\begin{aligned}
 \PoinS_{ M_m^{\mathrm{Dress}} }(t) &= 
\frac{\prod_{i=1}^r \left( 1-t^{d_i} \right)}{\prod_{j=1}^r \left( 1-t^{b_j} 
\right)} 
\qquad \Rightarrow \qquad 
\PoinS_{ M_m^{\mathrm{Dress}} }\left(\frac{1}{t}\right) 
= t^{-N}\PoinS_{ M_m^{\mathrm{Dress}} }(t)  \\
\for N&=\sum_{i=1}^r (d_i -b_i) = \#(\text{reflections in } \Wcal_\G ) 
-\#(\text{reflections in } \Wcal_{\Hh_m} ) \; .
\end{aligned}
\end{equation}
Here, we used the fact $\sum_{j=1}^r(b_j-1)=\#(\text{reflections in } 
\Wcal_{\Hh_m} )$, as stated earlier~\cite[Sec. 2]{Hanany:2016ezz}. 
At the same time, this proves that the order of $\PoinS_{ M_m^{\mathrm{Dress}} 
}(t)$ equals $N$, as given in the proposition.

In addition, we can relate the set of \Poincare\ series 
$\PoinS_{\mathfrak{J}(\hfrak_m)^{\Wcal_{\Hh_m}}}(t)$ to another fan in the Weyl 
chamber $\sigma$, the set of faces $\Fcal(\sigma)$.
\begin{myProp}
\label{prop:Dress_to_faces}
 The \Poincare\ series $P_{\G}(t;m)$ are constant along $\Relint(S_\tau)$ for 
all $\tau \in \Fcal(\sigma)$. In other words
\begin{equation}
 P_{\G}(t,m) \qquad \xleftrightarrow{\quad 1:1 \quad } \qquad \Relint(S_\tau)\, 
, \tau \in \Fcal(\sigma) \; .
\end{equation}
Hence, we define 
\begin{equation}
 \PoinS_{\Relint(S_\tau)} \coloneqq P_{\G}(t,m) \quad \text{for any } m\in 
\Relint(S_\tau) \; .
\end{equation}
\end{myProp}
The statement easily follows from recalling that the stabiliser $\Hh_m$ for 
$m\in \sigma$ depends only on the position of $m\in \Relint(\tau)$ for $\tau 
\in \Fcal(\sigma)$.
   % 
%%%%%%%%%%%%%%%%%%%%%%%%%%%%%%%%%%%%%%%%%%%%%%%%%%%%%%%%%%%%%%%%%%%%%%%%%%%%%%%%
   \section{Application to monopole formula}
\label{sec:application_monopole}
In this section we make contact between the algebraic set-up of 
Sec.~\ref{sec:monoid_algebras} and~\ref{sec:Casi_invariance}, and the monopole 
formula~\eqref{eqn:def_monopole_formula}. 
Recall the set-up introduced in~\cite{Hanany:2016ezz}, the matter content 
introduces a fan $F$ within the dominant Weyl chamber $\sigma$ of the GNO-dual 
group $\GNOG$ of a semi-simple gauge group $\G$. Moreover, the monopole formula 
intertwines two phenomena:
\begin{enumerate}[(A)]
 \item Hilbert series $\HilbS_{\Relint(S_\tau)}(t)$ for the canonical module 
$\K[\Relint(S_\tau)]$ for each cone $\tau\in F$.
\item \Poincare\ series $P_{\G}(t,m) = 
\PoinS_{\mathfrak{J}(\hfrak_m)^{\Wcal_{\Hh_m}}}(t)$ for the $\Hh_m$-invariant 
polynomials on 
$\hfrak_m \equiv \mathrm{Lie}(\Hh_m)$, where $\Hh_m = \mathrm{Stab}_m(G)$ 
denotes the residual gauge group.
\end{enumerate}
The monopole formula can then be rewritten as follows:

\begin{enumerate}[(1)]
 \item The lattice $\Lambda$ in question is the weight lattice of $\GNOG$. Let 
us emphasis this point again: the Coulomb branch is not fully specified by 
providing the gauge group and matter content (possibly in the form of a quiver 
diagram), but needs to be supplemented by the lattice of the magnetic 
charges\footnote{Choosing to quotient the gauge group by a discrete subgroup of 
its centre leads to different sub-lattices. This may lead to various possible 
choices of the magnetic weight lattice, in particular, when one quotients by 
common discrete subgroups of gauge group factors.}. 
This lattice follows from certain boundary conditions, which translate into 
the generalised Dirac quantisation condition for the set-up of this article.
  \item The grading employed varies for the different monoid and is 
given by the conformal dimension: for $m\in S_\tau $, 
$\tau \in F$ we use $\deg(m)\coloneqq \Delta(m)$. We note 
that this map is a monoid-homomorphism since it is a linear map on $S_\tau$ by 
construction.
Furthermore, this grading is indeed a positive grading by restriction to the 
class of good theories, defined by~\cite{Gaiotto:2008ak}.
  \item We can specialise the result~\eqref{eqn:HilbSeries_relint} for a monoid 
$S_\delta$ to the relevant grading. Collecting the cone generators in the 
set $X_\delta$, one obtains
\begin{equation}
 \HilbS_{\Relint(S_\delta)}(t) = \sum_{\delta \in \Delta'} \frac{\sum_{s\in 
\paraprim{\Lambda}{X_\delta}}  t^{\Delta(s)}}{\prod_{x \in X_\delta} 
\left(1-t^{\Delta(x)} \right)} 
= \sum_{\delta \in \Delta'} \frac{\sum_{k=1}^{d_\mathrm{max}(\delta)}  
\mathrm{card}(B_k)  t^{k}}{\prod_{x \in X_\delta} 
\left(1-t^{\Delta(x)} \right)} 
\label{eqn:HilbSeries_relint_special}
\end{equation}
with $d_\mathrm{max}(\delta) = \sum_{x \in X_\delta} \Delta(x) $.
  \item By Prop.~\ref{prop:Dress_to_faces}, the \Poincare\ series $P_\G(t;m)$ 
is constant along $\Relint(S_\delta)$ for the face $\delta\in \Fcal(\sigma)$ of 
the dominant Weyl chamber $\sigma$ with $m \in \Relint(S_\delta)$. 
The fan $F$ generated by the matter content is always a refinement of the fan 
$\Fcal(\sigma)$; therefore, the classical dressing factor 
is constant along each 
$\Relint(S_\tau)$ for each $\tau \in F$. Hence, we define for $\tau \in F$
\begin{equation}
\begin{aligned}
\PoinS_{\Relint(S_\tau)}(t) &\coloneqq \PoinS_{\Relint(S_\delta)}(t) \for 
\delta \in \Fcal(\sigma) \with \Relint(\tau)\subseteq \Relint(\delta) \; , \\
 \mathrm{or} \quad 
 \PoinS_{\Relint(S_\tau)}(t) &\coloneqq P_\G(t,m_\mathrm{int})
 \end{aligned}
\end{equation}
where $m_\mathrm{int}$ is any vector in $\Relint(S_\tau)$. One straightforward 
example can be constructed as follows: suppose $\tau = \cone(v_1,\ldots,v_s)$, 
and $\{v_1,\ldots,v_s\} \subseteq \Hilb(S_\tau)$ is the minimal set of cone 
generators, which is necessarily contained in the Hilbert basis. Then one 
can choose $m_\mathrm{int} = \sum_{i=1}^s v_i \in \Relint{S_\tau}$.
\item With the aforementioned comments, we can rewrite the monopole formula of 
a theory with gauge group $\G$ and matter content defining the fan $F$ as 
twisted Hilbert-\Poincare\ series
\begin{subequations}
\begin{align}
 \HS_{\G,F}(t)&= \sum_{\tau\in F} \PoinS_{\Relint(S_\tau)}(t) \cdot 
\HilbS_{\Relint(S_\tau)}(t)  
\label{eqn:rewrite_HS_init}\\
&= \PoinS_{\Relint(0)}(t)  \sum_{\tau \in F } 
\frac{\PoinS_{\Relint(S_\tau)}(t)}{\PoinS_{\Relint(0)}(t)} \cdot 
\HilbS_{\Relint(S_\tau)}(t)  \notag \\
&= \PoinS_{\Relint(0)}(t)  \sum_{\tau \in F } 
\PoinS_{M_\tau^{\text{Dress}}}(t) \cdot 
\HilbS_{\Relint(S_\tau)}(t) \;.
\end{align}%
\label{eqn:rewrite_HS}%
\end{subequations}%
\end{enumerate}
Formula~\eqref{eqn:rewrite_HS_init} provides a twisted sum over algebraic 
objects which are products of Cohen-Macaulay modules and Gorenstein rings. In 
some cases, for instance for rank one gauge groups, the monoid contributions 
are Gorenstein as well.
The observation strongly suggests
\begin{myConj}
\label{conj:Cohen_Macaulay}
 The Coulomb branch for $3$-dimensional $\Ncal =4$ gauge theories is 
Cohen-Macaulay.
\end{myConj}
Although we do not have a rigorous proof, evidence for this claim stems from the 
structures encoded in~\eqref{eqn:rewrite_HS_init} and from the absence of any 
counter-example, see also~\cite{Braverman:2016wma}.

One might wonder about the implications of Conj.~\ref{conj:Cohen_Macaulay}. From 
the computational viewpoint, 
supposing that the Coulomb branch is Cohen-Macaulay allows to apply 
Stanley's result~\cite[Thm.~4.4]{Stanley:1978} to check if the moduli space is 
even Gorenstein, purely from the Hilbert series. We emphasise that 
these identifications of the Coulomb branch being Cohen-Macaulay, Gorenstein (or 
even a complete intersection) are without explicitly knowing the associate 
coordinate ring.
Moreover, by Serre's criterion for normality~\cite[Thm.~2.2.22, 
p.~71]{Bruns:Herzog} 
one infers that the Coulomb branch would be a normal variety with a singular 
locus in $\mathrm{codim}\geq2$. Since the physical setting implies that the 
moduli space is a singular hyper-Kähler cone, the conjectured Cohen-Macaulay 
property would restrict the singularity\footnote{Another ``optimistical 
conjecture'' on the Coulomb branch has been put forward 
in~\cite{Braverman:2016wma}: it only has \emph{symplectic singularities}.}.
From the physics point of view, the aforementioned conjecture may shed light on 
$3$-dimensional dualities. For instance, the Higgs branch for a $3$-dimensional 
$\Ncal=4$ $\sprm(n)$ gauge theory with $2n$ flavours, as considered 
in~\cite{Ferlito:2016grh}, is a union of two cones, each of which is 
Cohen-Macaulay, but the union is not.
Then our claim in Conj.~\ref{conj:Cohen_Macaulay} on the Coulomb branch means 
that there cannot be a single quiver which reproduces this Higgs moduli space on 
the Coulomb branch side. In fact, for the Higgs branch example two quivers have 
been provided in~\cite{Ferlito:2016grh}, which generate each component. 
Therefore, mirror symmetry cannot be complete, as there exists a model 
where the mirror is not only unknown, but actually cannot exist if the above 
conjecture holds.

Starting from~\eqref{eqn:rewrite_HS} we proceed in several directions: firstly, 
utilising triangulations and simplicial refinements; secondly, working 
with free resolutions; and, lastly, combining triangulations and properties of 
the canonical module. The concepts will provide three different 
characterisations of the monopole formula. The first technique allows us to 
prove formal statements of the monopole formula, while the second method can 
be implemented in computer algebra software.
% 
%%%%%%%%%%%%%%%%%%%%%%%%%%%%%%%%%%%%%%%%%%%%%%%%%%%%%%%%%%%%%%%%%
%%%%%%%%%%%%%%%%%%%%%%%%%%%%%%%%%%%%%%%%%%%%%%%%%%%%%%%%%%%%%%%%%
%
\subsection{Simplicial refinements}
From~\eqref{eqn:HilbSeries_relint_special} we know how the Hilbert series for 
the relative interior of a monoid is computed via a triangulation. To apply 
this technique to the entire monopole formula~\eqref{eqn:rewrite_HS} we, roughly 
speaking, extend the concept of triangulation to the fan $F$. To be more 
precise, we need to consider a \emph{simplicial refinement} 
$\Phi$ of the fan $F$, which is a fan comprised solely of simplicial cones $\tau 
\in \Phi$ such that for any cone $\delta\in \Phi$ there exists at least one cone 
$C \in F$ such that $\delta \subseteq C$. The existence of such a refinement 
follows by induction over the top-dimensional cones of $F$ and their simplicial 
triangulations, which one chooses to be mutually compatible.

As a consequence, the refined fan $\Phi$ is a disjoint union of the 
relative interior of simplicial cones and, similarly to 
Lem.~\ref{lem:prefered_triangle}, one can choose a refinement such that all 
simplicial cones are spanned by the cone generators of the cones in $F$.
This then allows us to write~\eqref{eqn:rewrite_HS} as
\begin{subequations}
\begin{align}
  \HS_{\G,F}(t) = 
  \PoinS_{\Relint(0)}(t)  \sum_{\tau \in \Phi } 
\PoinS_{M_\tau^{\text{Dress}}}(t) \cdot \HilbS_{\Relint(S_\tau)}(t) \; ,
\end{align}
but as each cone $\tau\in \Phi$ is simplicial, we can 
apply~\eqref{eqn:HilbSeries_relint_special} and obtain
\begin{align}
   \HS_{\G,F}(t) &= 
  \PoinS_{\Relint(0)}(t)  \sum_{\delta \in \Phi } 
\PoinS_{M_\delta^{\text{Dress}}}(t) \cdot 
\frac{\sum_{k=1}^{d_\mathrm{max}(\delta)}  
\mathrm{card}(B_k)  t^{k}}{\prod_{x \in X_\delta} 
\left(1-t^{\Delta(x)} \right)}  
\equiv \frac{R_{\G,F}(t)}{P_{\G,F}(t)} \; ,\\
R_{\G,F}(t) &= 
\sum_{\delta \in \Phi }  \left\{  
\PoinS_{M_\delta^{\text{Dress}}}(t) 
\cdot 
\left[\left( \sum_{k=1}^{d_\mathrm{max}(\delta)}  
\mathrm{card}(B_k)  t^{k} \right)
\cdot
\prod_{x \in X\setminus X_\delta} \left(1-t^{\Delta(x)} \right) \right]  
\right\} \; , 
\label{eqn:HS_refinement_numerator}\\
P_{\G,F}(t) &= 
\frac{\prod_{y \in X} \left(1-t^{\Delta(y)} \right) }{\PoinS_{\Relint(0)}(t)} 
 =\prod_{i=1}^r \left(1-t^{d_j} \right) \prod_{y \in X} \left(1-t^{\Delta(y)} 
\right) \;. \label{eqn:HS_refinement_denominator}
\end{align}%
\label{eqn:rewrite_HS_refinement}%
\end{subequations}%
We observe that the numerator~\eqref{eqn:HS_refinement_numerator} is a 
polynomial in $t$ with integer 
coefficients, and the denominator~\eqref{eqn:HS_refinement_denominator} is 
comprised of the Casimir invariants (of 
degrees $d_j$) of the gauge group $\G$ together with the set $X$ of cone 
generators of $F$.
% 
%%%%%%%%%%%%%%%%%%%%%%%%%%%%%%%%%%%%%%%%%%%%%%%%%%%%%%%%%%%%%%%%%
%%%%%%%%%%%%%%%%%%%%%%%%%%%%%%%%%%%%%%%%%%%%%%%%%%%%%%%%%%%%%%%%%
%
\subsection{Free resolutions}
A rather different viewpoint is taken by the use of free resolutions as 
introduced in Sec.~\ref{subsec:resolutions}.
What we are after are the relations or syzygyies between the minimal generators 
in a given Hilbert basis. In order to obtain their relations, based on the 
position in the monoid $S$ itself, we need to employ a slightly different 
grading at first. Again, the grading varies for the different monoids, but 
is now given by the graph of the conformal dimension that is for $m\in S_\tau 
$, $\tau \in F$, we use $\deg(m)\coloneqq \Gamma_\Delta(m) =(m,\Delta(m))$. We 
note that this map is a monoid-homomorphism since it is the identity in the 
first factor and a linear map in the second factor. Furthermore, the grading 
is positive on the class of good theories, because the conformal 
dimension ensures that $\Gamma_\Delta(m) =(0,0)$ has the unique 
solution $m=0$. 
 
Given this grading and having determined the Hilbert basis for each monoid 
$S_\tau$, $\tau \in F$, we can simply apply algorithms from algebraic 
geometry~\cite{Miller:2005,Macaulay2book} to compute the corresponding lattice 
ideals. As a consequence, the contribution of a monoid $S_\tau$ takes the form 
\begin{equation}
 \HilbS_{S_\tau} (\bt) = \frac{\Kcal(S_\tau,\bt)}{\prod_{x \in 
\Hilb(S_\tau)}\left(1-\bt^{\Gamma_\Delta(x)} \right) } \; .
\label{eqn:HilbS_resolution}
\end{equation}
From the previous construction~\eqref{eqn:rewrite_HS}, we know that we are 
forced to sum over the relative interiors of the monoids. As we outlined 
in~\cite{Hanany:2016ezz}, one can realise this by means of the 
\emph{exclusion-inclusion principle} as follows
\begin{equation}
 \HilbS_{\Relint(S_\tau)}(\bt) = \HilbS_{S_\tau} (\bt) + 
\sum_{\delta\in\Fcal(\tau)\setminus\{\tau\}} (-1)^{\codim(\delta,\tau)} 
\HilbS_{S_\delta} (\bt) 
= \sum_{\delta\in\Fcal(\tau)} (-1)^{\codim(\delta,\tau)} 
\HilbS_{S_\delta} (\bt) \; ,
\label{eqn:exclusion_inclusion}
\end{equation}
where we defined $\codim(\delta,\tau)\coloneqq \dim(\tau) - \dim(\delta)$, and
$\Fcal(\tau)$ denotes the set of faces of $\tau$. One 
applies~\eqref{eqn:HilbS_resolution} to each summand and obtains
\begin{subequations}
\begin{align}
 \HilbS_{\Relint(S_\tau)}(\bt) &=
 \sum_{\delta\in\Fcal(\tau)} (-1)^{\codim(\delta,\tau)} 
\frac{\Kcal(S_\delta,\bt)}{\prod_{x \in 
\Hilb(S_\delta)}\left(1-\bt^{\Gamma_\Delta(x)} \right) } \\*
&=\frac{\sum_{\delta\in\Fcal(\tau)} 
\left\{ (-1)^{\codim(\delta,\tau)} \cdot \Kcal(S_\delta,\bt) \cdot
\prod_{x \in 
 \Hilb(S_\tau)\setminus\Hilb(S_\delta)} \left(1-\bt^{\Gamma_\Delta(x)} \right) 
\right\}
}{ \prod_{y \in 
\Hilb(S_\tau)}\left(1-\bt^{\Gamma_\Delta(y)} \right)
} \; .
\end{align}
\end{subequations}
However, the multi-fugacity $\bt$ is too much in most 
cases. For instance, if there are no global symmetries we only employ the 
$\su_R$-isospin grading. The reason why we need the monoid 
$\Gamma_\Delta(S_\tau)$ is to compute the lattice ideals $I_L$ correctly, 
because 
the syzygyies between the minimal generators are determined by the information 
of $S_\tau$ and not only the conformal dimension. 

For a rank $r$ gauge group, $m$ is an $r$-dimensional vector and the 
multi-fugacity $\bt$ consists of $r+1$ individual fugacities $t_j$, for 
$j=1,\ldots, r, r+1$. To reduce to the physically desired scenario we define 
\begin{equation}
 \HilbS_{\Relint(S_\tau)} (t) \coloneqq \lim_{\substack{t_i \to 1 \\ i 
=1,\ldots, r}} \HilbS_{\Relint(S_\tau)} (t_1,\ldots,t_r,t_{r+1}\equiv t)
\label{eqn:HS_def_limit}
\end{equation}
With this definition we are well-equipped to provide the full expression of the 
monopole formula 
\begin{subequations}
\begin{align}
 \HS_{\G,F}(t)&= \sum_{\tau\in F} \PoinS_{\Relint(S_\tau)}(t) \cdot
\sum_{\delta\in\Fcal(\tau)} (-1)^{\codim(\delta,\tau)} 
\frac{\Kcal(S_\delta,t)}{\prod_{x \in 
\Hilb(S_\delta)}\left(1-t^{\Delta(x)} \right) }
\equiv \frac{\widetilde{R}_{\G,F}(t)}{\widetilde{P}_{\G,F}(t)} ,  \\
\widetilde{R}_{\G,F}(t) &= 
\sum_{\tau \in F } \PoinS_{M_\tau^{\text{Dress}}}(t) \cdot 
\sum_{\delta\in\Fcal(\tau)}
\left\{ (-1)^{\codim(\delta,\tau)} \cdot \Kcal(S_\delta,t) 
\prod_{x \in 
 \Hilb(F)\setminus\Hilb(S_\delta)} \left(1-t^{\Delta(x)} \right) 
\right\} , \label{eqn:HS_resolution_numerator}\\
\widetilde{P}_{\G,F}(t) &=\prod_{i=1}^r \left(1-t^{d_j} \right) \cdot 
\prod_{y \in \Hilb(F)}\left(1-t^{\Delta(y)} \right)   ,
\label{eqn:HS_resolution_denominator}
\end{align}%
\label{eqn:rewrite_HS_resolution}%
\end{subequations}%
where $\Hilb(F) \coloneqq \cup_{\tau\in F} \Hilb(S_\tau)$ represents the 
collection of all minimal generators of the various Hilbert bases. In contrast 
to~\eqref{eqn:rewrite_HS_refinement}, the 
denominator~\eqref{eqn:HS_resolution_denominator} is comprised of the Casimir 
invariants together with all monoid generators in the fan $F$. The 
numerator~\eqref{eqn:HS_resolution_numerator} is again a 
polynomial in $t$ with integer coefficients.
% 
%%%%%%%%%%%%%%%%%%%%%%%%%%%%%%%%%%%%%%%%%%%%%%%%%%%%%%%%%%%%%%%%%%%%%%%%
%%%%%%%%%%%%%%%%%%%%%%%%%%%%%%%%%%%%%%%%%%%%%%%%%%%%%%%%%%%%%%%%%%%%%%%%
%
\subsection{Canonical module}
\label{subsec:use_canonical_module}
For completeness, we provide a third approach, which combines triangulations 
and the properties of canonical modules.

Suppose $\tau\in F$ with $S_{\tau}=\tau \cap\Lambda$, and let $\Delta$ be a 
triangulation of $\tau$. Then by combining Lem.~\ref{lem:properties_triangle}, 
\ref{lem:prefered_triangle}, and ~\ref{lem:Stanley}, one deduces
\begin{equation}
\begin{aligned}
 \HilbS_{\Relint(S_{\tau})}(t) 
 &= (-1)^{\rank(S_{\tau})} \HilbS_{S_{\tau}}(t^{-1})
 = (-1)^{\rank(S_{\tau})} \sum_{\delta \in \Delta} 
\HilbS_{\Relint(S_\delta)}(t^{-1}) \\
 &=  \sum_{\delta \in \Delta} (-1)^{\rank(S_\tau) - \rank(S_\delta)} 
\HilbS_{S_\delta}(t) 
=  \sum_{\delta \in \Delta} (-1)^{\dim(\tau) - \dim(\delta)} 
\HilbS_{S_\delta}(t) \; ,
\end{aligned}
\end{equation}
which is a result already known by Stanley~\cite{Bruns:2009}. Note that the 
Hilbert series $\HilbS_{S_\delta}(t)$ for a monoid ring $\K[S_\delta]$ is given 
by~\eqref{eqn:HS_monoid}. Employing this for the monopole formula yields
\begin{subequations}
\begin{align}
 \HS_{\G,F}(t) &= \sum_{\tau \in F} \PoinS_{\Relint(S_\tau)}(t) \cdot 
\sum_{\delta \in \Delta(\tau)} (-1)^{\codim(\delta,\tau)} 
\HilbS_{S_\delta}(t) \equiv \frac{\widehat{R}_{\G,F}(t)}{\widehat{P}_{\G,F}(t)} 
\; , \\
\widehat{R}_{\G,F}(t) &= \sum_{\tau \in F} 
\PoinS_{M_{\Relint(\tau)}^{\mathrm{Dress}}}(t) 
\cdot \sum_{\delta \in \Delta(\tau)} \left\{ (-1)^{\codim(\delta,\tau)} \cdot
Q_{\delta}(t) \prod_{y\in X\setminus X_{\delta}} \left( 1-t^{\Delta(y)} \right) 
\right\} \; , \\
\widehat{P}_{\G,F}(t) &=\prod_{i=1}^r \left( 1-t^{d_i} \right) \prod_{x \in X} 
\left(1-t^{\Delta(x)} \right) \; , \label{eqn:HS_canonical_denominator}
\end{align}
\label{eqn:rewrite_HS_canonical-module}
\end{subequations}
where $X_\delta$ denotes the cone generators of $\delta$, and $X$ the 
collection of all cone generators of the matter fan $F$.

Let us compare the three representations~\eqref{eqn:rewrite_HS_refinement}, 
\eqref{eqn:rewrite_HS_resolution}, \eqref{eqn:rewrite_HS_canonical-module}. 
Formulae~\eqref{eqn:rewrite_HS_resolution} and 
\eqref{eqn:rewrite_HS_canonical-module} are exclusion-inclusion type of 
alternating sums. Moreover, the appearing numerators $Q$ and $\mathcal{K}$ 
are obtained from free resolutions; therefore, are integer polynomials with 
potentially negative signs. In contrast, 
formula~\eqref{eqn:rewrite_HS_refinement} contains positive summands and all 
numerators are non-negative integer valued. Hence, this expression is most 
suitable for the analysis of the pole structure of 
Sec.~\ref{sec:pole_structure}.

However, the results~\eqref{eqn:rewrite_HS_refinement}, 
\eqref{eqn:rewrite_HS_canonical-module} employ triangulations or simplicial 
refinements for a further subdivision of the matter fan. Additionally, the 
Hilbert series for the individual monoids still have to be computed by either 
free resolutions of the lattice ideals or some counting algorithm for the 
number of points in fundamental parallelepiped. Whereas, 
formula~\eqref{eqn:rewrite_HS_resolution} just requires the combinatorial data 
of the matter fan plus the free resolutions of each lattice ideal. Thus, making 
this approach well-suited for explicit calculations, such as present in 
Sec.~\ref{sec:implementation}.

Lastly, formula~\eqref{eqn:rewrite_HS_canonical-module} is a mixture between 
the combinatorial and the algebraic approach. Its main feature is that it is a 
twisted sum of \Poincare\ series of normal Gorenstein rings and Hilbert series 
of normal Cohen-Macaulay monoid rings. The latter can in some instances be 
Gorenstein, too. Compared to~\eqref{eqn:rewrite_HS_resolution}, the 
denominator~\eqref{eqn:HS_canonical_denominator} is determined by the Casimir 
invariants of $\G$ and the cone generators of $F$ in contrast to the union of 
all Hilbert bases.
% 
%%%%%%%%%%%%%%%%%%%%%%%%%%%%%%%%%%%%%%%%%%%%%%%%%%%%%%%%%%%%%%%%%%%%%%%%
%%%%%%%%%%%%%%%%%%%%%%%%%%%%%%%%%%%%%%%%%%%%%%%%%%%%%%%%%%%%%%%%%%%%%%%%
%
\subsection{Comments on matter fan}
\label{subsec:fan}
Now we comment on our construction of the matter fan described in the 
introduction and introduced in~\cite{Hanany:2016ezz}. 
First of all, let us remind ourselves why we restrict to relevant weights only. 
Suppose $\mu$ is a weight vector of $\G$ such that $\mu \in \sigma^\vee \cap 
\Lambda_w(\G)$, where $\sigma^\vee$ is the dual cone of the dominant Weyl 
chamber $\sigma$ of the GNO-dual group $\GNOG$. Then $\sigma^\vee = 
\cone(\alpha_j | j=1,\ldots,\rank(\G)) $ with $\alpha_j $ are the simple roots 
of $\G$. By definition, $\sigma \cap H_{\mu} \eqqcolon \tau_\mu$ defines the 
face $\tau_\mu$ of $\sigma$. Hence, $\sigma \cap H_{\mu}^+ = \sigma$ and 
$\sigma \cap H_{\mu}^- = \tau_\mu$. 
Similarly, if $\mu$ is a weight vector for which $\mu \in (-\sigma^\vee) \cap 
\Lambda$, we see that $\cone(\mu,\{\alpha_j | j=1,\ldots,\rank(\G) \})$ is not a
strongly convex, because it contains the real line $\R \, \mu$. 
Hence, the dual cone $\cone(\mu,\{\alpha_j | j=1,\ldots,\rank(\G) \})^\vee = 
\bigcap_{j=1,\ldots, \rank(\G)} H_{\alpha_j}^+ \cap H_{\mu}^+$ is not of 
maximal dimension, and since $\mu$ and $-\mu$ define the same hyper-plane, 
$\mu$ gives rise to the same face as $-\mu$.
Therefore, the absolute value 
$|\mu(m)|$ is already resolved on the entire Weyl chamber $\sigma$ for weights 
satisfying $\pm \mu \in \sigma^\vee \cap \Lambda_w(\G)$.

For the relevant weights, we introduced the cones~\eqref{eqn:cones_cap_Weyl}. 
In order the define the matter fan it is sufficient to work with the 
top-dimensional cones, because all other cones are generated by the properties 
of a fan.
Here, we identify the set of cones that are top-dimensional and thus generate 
the entire matter fan. 
For simplicity, start with two relevant weights $\mu_1, \mu_2 \in 
\Lambda_w(\G)$. Consider the cones
\begin{equation}
\begin{aligned}
 \tau_{\epsilon_1,\epsilon_2} 
 &= \bigcap_{j = 1, \ldots, \rank(\G)} 
 H_{\alpha_j}^+ \cap H_{\mu_1}^{\epsilon_1} \cap H_{\mu_2}^{\epsilon_2} 
= \bigcap_{j =1, \ldots, \rank(\G)} H_{\alpha_j}^+ \cap H_{\epsilon_1 \mu_1}^+ 
\cap H_{\epsilon_2 \mu_2}^+ \\
&= \cone \left(\epsilon_1 \mu_1, \epsilon_2 \mu_2, \{\alpha_j | j = 1,\ldots, 
\rank(\G)\} \right)^\vee \with \epsilon_i = \pm \; .
\end{aligned}
\end{equation}
We distinguish the following three cases (for fixed $\epsilon_1$):
\begin{enumerate}[(i)]
 \item For $\mu_2 \in \cone \left(\epsilon_1 \mu_1, \{\alpha_j 
| j = 1,\ldots, \rank(\G)\} \right)$, $\mu_2$ generates a face of 
$\tau_{\epsilon_1}$, and does not further subdivide the cone 
$\tau_{\epsilon_1}$.
 \item For $\mu_2 \in -\cone \left(\epsilon_1 \mu_1, \{\alpha_j 
| j = 1,\ldots, \rank(\G)\} \right)$, $\cone \left(\epsilon_1 \mu_1, 
\mu_2, \{\alpha_j | j = 1,\ldots, 
\rank(\G)\} \right)$ contains the real line $\R \, \mu_2$. As such the dual 
cone is not top-dimensional, and previous arguments show that the dual cone is 
a proper face of $\sigma$ or the trivial cone.
  \item For $\pm \mu_2 \notin \cone \left(\epsilon_1 \mu_1, \{\alpha_j 
| j = 1,\ldots, \rank(\G)\} \right)$, the two cones 
$\tau_{\epsilon_1,\epsilon_2} $ lead to a division of $\tau_{\epsilon_1}$ into 
two  top-dimensional, strongly convex sub-cones.
\end{enumerate}
For the generic case, the cones 
\begin{equation}
 \tau_{\epsilon_1,\ldots,\epsilon_{|\Gamma|}} = 
 \bigcap_{i=1,\ldots,|\Gamma|} H_{\epsilon_i \mu_i}^+ \cap  \sigma  
 = \cone\left(\{\epsilon_i \mu_i \,| \; i =1,\ldots, |\Gamma|\},\{\alpha_j \,| 
\;  j=1,\ldots, \rank(G)\}\right)^\vee
\end{equation}
of~\eqref{eqn:cones_cap_Weyl} are maximal dimensional 
whenever $\{\epsilon_i \mu_i \,| \; i =1,\ldots, |\Gamma|\}$ is a necessary 
subset of the cone generators for the dual cone of 
$\tau_{\epsilon_1,\ldots,\epsilon_{|\Gamma|}}$. Or in other words,
\begin{equation}
\forall k\in 1,\ldots,|\Gamma|\; : \qquad  \pm \mu_k \notin 
\cone\left(\{\epsilon_i \mu_i \,| \; i \neq k\},\{\alpha_j \,| \;  j=1,\ldots, 
\rank(G)\}\right) \; .
\end{equation}
The resulting set of relevant, top-dimensional cones suffices to generate 
the fan $F$ via its defining properties: each faces of a cone in $F$ is a cone 
in the fan, and the intersection of two cone is a face of each.
% 
%%%%%%%%%%%%%%%%%%%%%%%%%%%%%%%%%%%%%%%%%%%%%%%%%%%%%%%%%%%%%%%%%%%%%%%%
%%%%%%%%%%%%%%%%%%%%%%%%%%%%%%%%%%%%%%%%%%%%%%%%%%%%%%%%%%%%%%%%%%%%%%%%
%
\subsection{Product gauge groups}
\label{subsec:product_groups}
So far we considered a semi-simple gauge group $\G$ and, hence, a semi-simple 
$\GNOG$. For physically interesting theories we need to consider product 
groups of semi-simple $\G_i$ and abelian $\uo$ factors
\begin{equation}
 \prod_{i} \G_i \times \prod_a \uo_a \; .
 \label{eqn:prod_gauge_group}
\end{equation} 

Semi-simple gauge groups $\G$ are nice, because it is a priori clear that the 
fan $F$ consists of strongly convex cones only. Thus, the monoids are positive 
and the Hilbert basis is meaningful. The reason is that the dominant Weyl 
chamber $\sigma$ of $\GNOG$ is by itself a strongly convex cone and the set of 
faces $\Fcal(\sigma)$ is understood as fan with strongly convex cones only. 
Since the fan $F$ is a refinement of $\Fcal(\sigma)$, all cones $\tau \in F$ 
are strongly convex as well.

For the product $\G_1\times \G_2$ of two semi-simple groups, the GNO-dual group 
is $\GNOG_1\times \GNOG_2$, by~\cite{Goddard:1976qe}. Then dominant Weyl 
chambers $\sigma_1 \subset \R^{\rank(\G_i)}$ combine via the Minkowski sum into 
a polyhedral rational cone $\sigma_1 + \sigma_2 \subset 
\R^{\rank(\G_1)+\rank(G_2)}$. Moreover, since each $\sigma_i$ is strongly 
convex, also the sum is a strongly convex cone. Therefore, the fan $F$ is again 
a refinement of the set of faces of a strongly convex cone and, consequently, 
all cones in $F$ are strongly convex.

In the case of a product $\uo \times \G$, for a semi-simple $\G$, the dual 
group is still considered as $\uo \times \GNOG$. The crux, however, lies in the 
appropriate notion of the dominant Weyl chamber. Since the magnetic weight 
lattice of $\uo$ is $\Z$, we associate $\R$ as Weyl chamber. But then $\R + 
\sigma\subset \R^{\rank(\G)+1}$ as combined Weyl chamber is not strongly 
convex any more. Fortunately, obtaining a fan $F$ containing solely strongly 
convex cones is achievable as long as there is any hyper-plane $H_\mu$ that 
intersects the $\R$ part non-trivial. This is realised in all reasonable 
scenarios with a contribution $|\mu(m)| \propto |a\cdot m_1 + \ldots|$ to the 
conformal dimension, wherein $m_1$ is the magnetic charge in $\uo$ direction. 
In other words, the hyper-multiplets have to couple to the $\uo$, otherwise it 
would be a decoupled factor anyway. With this (mild) assumption on the 
hyper-plane arrangement defining $F$, the conclusion is as before: all 
cones in $F$ are strongly convex.

In summary, the considerations of monoid algebras, Hilbert series thereof, and 
applications to the monopole formula are valid for gauge groups of the 
form~\eqref{eqn:prod_gauge_group}. Note that the consequences of a product 
gauge group for the \Poincare\ series for the Casimir 
invariance for such product gauge groups has already been discussed 
in~\cite{Hanany:2016ezz}.
% %%%%%%%%%%%%%%%%%%%%%%%%%%%%%%%%%%%%%%%%%%%%%%%%%%%%%%%%%%%%%%%%%%%%%%%%%%%%%%%%
  \section{Pole structure of the monopole formula}
\label{sec:pole_structure}
Having established the generalities in Sec.~\ref{sec:monoid_algebras} 
and~\ref{sec:Casi_invariance}, we now prove certain statements about the 
monopole formula that have implicitly been used in earlier 
works~\cite{Cremonesi:2013lqa,Hanany:2016ezz}.
\subsection{Order of pole at \texorpdfstring{$t=1$}{t=1}}
Suppose we have a theory with a rank $r$ gauge group $\G$ and 
matter content defining the fan $F$. Then we claim the following:
\begin{myProp}
\label{prop:pole_t=1}
 The order of the pole of the monopole formula at $t=1$ equals twice 
the rank of the gauge group and, thus, coincides with the complex 
dimension of the moduli space.
\end{myProp}
The starting point for the proof is the monopole 
formula~\eqref{eqn:rewrite_HS_refinement}, which is expressed for a simplicial 
refinement $\Phi$ of the fan $F$. As stated in Prop.~\ref{prop:Casimir_inv}, 
$\PoinS_{M_{\tau}^{\text{Dress}}}(t)$ is a finite polynomial with 
non-negative integer 
coefficients, i.e.\ a polynomial without any zeros or poles at $t=1$.

For each simplicial cone $\tau \in \Phi$, the Hilbert 
series of the relative interior of the corresponding monoid $S_\tau$ has the 
form~\eqref{eqn:HilbSeries_relint_special}.
We observe that the Hilbert series~\eqref{eqn:HilbSeries_relint_special} has a 
polynomial with non-negative integer coefficients as 
numerator, meaning the numerator has no zero at $t=1$. Thus, the rational 
function $\HilbS_{\Relint(S_\tau)}(t)$ has a pole of order $\dim(\tau)$ at 
$t=1$.

Putting all the pieces together in~\eqref{eqn:rewrite_HS_refinement} we arrive 
at
\begin{align}
  \HS_{\G,F}(t) &= \underbrace{\PoinS_{\Relint(0)}(t)}_{\text{pole 
of order $r$}} 
 \; \sum_{\tau \in \Phi} 
\underbrace{\PoinS_{M_{\tau}^{\text{Dress}}}(t)}_{\text{
no poles or zeros}} 
\cdot \underbrace{\HilbS_{\Relint(S_\tau)}(t)}_{\text{pole of order 
$\dim(\tau)$}} \;.
\label{eqn:pole_structure}
\end{align}
From which we infer that the Hilbert series has a pole of order $2r$ at $t=1$, 
as the maximal dimensional cones have dimension equal $r$.
This argument relies on the existence 
of a top-dimensional cone in a simplicial refinement $\Phi$. This holds, 
because $\Phi$ is a refinement of $F$ and, hence, also a refinement of 
$\Fcal(\sigma)$. Since $\Fcal(\sigma)$ contains $\sigma$ as the only 
$r$-dimensional cone, it follows that $F$ as well as $\Phi$ have at least one 
$r$-dimensional cone.
% 
%%%%%%%%%%%%%%%%%%%%%%%%%%%%%%%%%%%%%%%%%%%%%%%%%%%%%%%%%%%%%%%%%%%%%%%%%%%
%%%%%%%%%%%%%%%%%%%%%%%%%%%%%%%%%%%%%%%%%%%%%%%%%%%%%%%%%%%%%%%%%%%%%%%%%%%
%
\subsection{Order of pole at \texorpdfstring{$t\to +\infty$}{t->oo}}
As before, consider a rank $r$ gauge group $\G$ and 
matter content defining the fan $F$, then we prove the following statement for 
the monopole formula:
\begin{myProp}
\label{prop:pole_t=infty}
The order of the pole at $t\to +\infty$ equals the rank 
of the gauge group, provided the fugacity is $t$. For fugacity $t^2$, the 
order of the pole at infinity is twice the rank of the gauge group.
\end{myProp}
Again, we start from~\eqref{eqn:rewrite_HS_refinement} and the notation  
$X=\{v_1,\ldots,v_r\}$ for the set of all cone generators in $F$.
Next, the order of the pole of $\HS_{\G,F}(t)$ at $t\to \infty$ 
equals the 
difference in maximal degrees of denominator $P_{\G,F}(t)$ and numerator 
$R_{\G,F}(t)$. We infer
\begin{align}
 \max \deg (P_{\G,F}) = \sum_{j=1}^r d_j + \sum_{x \in X} \Delta(x) \; ,
\end{align}
where $d_j$ are the degrees of the Casimir invariants of $\G$.
In addition, for each summand in the numerator $R_{\G,F}(t)$ we obtain
\begin{align}
 \max \deg \left[
\left( \sum_{k=1}^{d_{\mathrm{max}}(\tau) }  (\mathrm{card}(B_k)) \cdot t^k 
\right)
 \prod_{y\in X \setminus X_\tau} (1-t^{\Delta(y)}) \right] 
 &= d_{\mathrm{max}}(\tau) + \sum_{y\in X \setminus X_\tau} \Delta(y)  \\*
&= \sum_{x\in X_\tau} \Delta(x) + \sum_{y\in X \setminus X_\tau} \Delta(y) = 
\sum_{x\in X} \Delta(x) \; , \notag
\end{align}
which implies that the term in square brackets 
of~\eqref{eqn:rewrite_HS_refinement} has always the same maximal degree, 
independent of the cone $\tau \in \Phi$. Then, all left to do is 
to find the cone with maximal degree polynomial 
$\PoinS_{M_{\tau}^{\text{Dress}}}=\frac{P_{\Relint(S_\tau)}(t)}{P_{\Relint(0)}
(t) } $, but this is clearly the 
case when the residual gauge group of $\Relint(S_\tau)$ is just the maximal 
torus of $\G$. 
In detail, using the results from Prop.~\ref{prop:Casimir_inv} or~\cite[App. 
A]{Hanany:2016ezz} we find
\begin{align}
 \max \deg \frac{P_{\G}(\Relint(S_\tau))}{P_{\G}(0)} = \max \deg \prod_{i=1}^r 
\sum_{l_i}^{d_i-1} t^{l_i} = \max \deg \prod_{i=1}^r t^{d_i -1} = \sum_{i=1}^r 
(d_i -1) \; ,
\end{align}
% pole
and finally
\begin{align}
 \max \deg (R_{\G,F}) &= \sum_{i=1}^r (d_i -1) +\sum_{x\in X} \Delta(x) \;, \\
 \max \deg (P_{\G,F})- \max \deg (R_{\G,F}) &= \sum_{i=1}^r 1 = r \; .
\end{align}
Hence, the claim holds: the order of the pole at $t \to \infty$ equals the rank 
of the gauge group for fugacity $t$. Furthermore, using 
$t^2$ fugacity instead of $t$ results in overall factors of $2$ in the maximal 
degrees of numerator and denominator. Therefore, the order of the pole is 
altered to $2r$ in this case.
% 
%%%%%%%%%%%%%%%%%%%%%%%%%%%%%%%%%%%%%%%%%%%%%%%%%%%%%%%%%%%%%%%%%%%%%%%%%%%%%%%
%%%%%%%%%%%%%%%%%%%%%%%%%%%%%%%%%%%%%%%%%%%%%%%%%%%%%%%%%%%%%%%%%%%%%%%%%%%%%%%
% 
\subsection{Order of pole at \texorpdfstring{$t=-1$}{t=-1}}
Now, we turn to the less clear poles: to start with, let us consider $t=-1$. As 
before, consider a gauge group $\G$ of rank $r$ together with a fan 
$F$. Denote the collection of cone generators of $F$ by $X$ and let 
$\{b_j(\tau)\}_{j=1,\ldots,r}$ denote the degree of the Casimir invariants for 
$\Hh_{m}$ for any $m\in \Relint(S_\tau)$.
\begin{myProp}
\label{prop:pole_t=-1}
 The order of the pole of the Hilbert series at $t=-1$ satisfies
 \begin{align}
  \text{order of pole}(\HS_{\G,F}(t))|_{t=-1} &\leq \max_{\tau \in F}\left\{ 
\#\left\{ b_j(\tau) \, \big| \; b_j(\tau)=\mathrm{even} \, \right\} \right\} 
+ \#\left\{ x \in X \, \big| \, \Delta(x)=\mathrm{even} \right\} \notag \\
&\leq r + \#\left\{ x \in X \, \big| \, \Delta(x)=\mathrm{even} \right\} \; ,
\end{align}
\end{myProp}
To prove this upper bound, we use the explicit form of~\eqref{eqn:rewrite_HS} 
for a simplicial refinement $\Phi$ of the fan $F$ as follows:
\begin{align}
\label{eqn:HS_pole_calc}
  \HS_{\G,F}(t)&= \sum_{\tau\in \Phi} \frac{1}{\prod_{j=1}^r 
\left(1-t^{b_j(\tau)} \right)} \cdot 
\frac{\sum_{k=1}^{d_\mathrm{max}(\tau)}  
\mathrm{card}(B_k)  t^{k}}{\prod_{x \in X_\tau} 
\left(1-t^{\Delta(x)} \right)} \; .
\end{align}
The order of the pole for the contribution of each cone $\tau$ can be estimated 
as 
\begin{align} 
Z_\tau\coloneqq \#\left\{ b_j(\tau) \, \big| \; b_j(\tau)=\mathrm{even} \, 
\right\} 
&+ \#\left\{ x \in X_\tau \, \big| \, \Delta(x)=\mathrm{even} \right\} \\*
&- \text{order of zero} \left\{ \sum_{k=1}^{d_\mathrm{max}(\tau)}  
\mathrm{card}(B_k)  t^{k} \, \big| \, t=-1 \right\} \notag \; .
\end{align}
For each piece we have rough bounds
\begin{subequations}%
\begin{align}%
 0 &\leq \#\left\{ b_j(\tau) \, \big| \; b_j(\tau)=\mathrm{even} \, 
\right\} \leq r \; , \\
0 &\leq \#\left\{ x \in X_\tau \, \big| \, \Delta(x)=\mathrm{even} \right\} 
\leq \#\left\{ x \in X \, \big| \, \Delta(x)=\mathrm{even} \right\} \; , \\
0&\leq \text{order of zero} \left\{ \sum_{k=1}^{d_\mathrm{max}(\tau)}  
\mathrm{card}(B_k)  t^{k} \, \big| \, t=-1 \right\} \leq d_\mathrm{max}(\tau) 
-1 \; .
\end{align}%
\label{eqn:estimates}%
\end{subequations}%
Neglecting cancellations between the summands in~\eqref{eqn:HS_pole_calc} we 
estimate an the upper bound on the order of the pole as 
\begin{align}
 \text{order of pole}(\HS_{\G,F}(t))|_{t=-1} \leq \max_{\tau \in\Phi} Z_\tau \; 
,
\end{align}
after using~\eqref{eqn:estimates} we obtain the claim.
% 
%%%%%%%%%%%%%%%%%%%%%%%%%%%%%%%%%%%%%%%%%%%%%%%%%%%%%%%%%%%%
%%%%%%%%%%%%%%%%%%%%%%%%%%%%%%%%%%%%%%%%%%%%%%%%%%%%%%%%%%%%
%
\subsection{Order of pole at other roots of unity}
Lastly, we consider other roots of unity of the form $t=e^{2\pi 
\im \frac{q}{p}}$, where we can restrict ourselves to co-prime integers $q$, 
$p$ with $0< q < p $. Consider a gauge group $\G$ of rank $r$ 
together with a fan $F$. Denote the collection of cone generators of $F$ by $X$ 
and let $\{b_j(\tau)\}_{j=1,\ldots,r}$ denote the degree of the Casimir 
invariants for $\Hh_{m}$ for any $m\in S_\tau$.
\begin{myProp}
\label{prop:pole_t=root_of_unity}
 The order of the pole of the Hilbert series at $t=e^{2\pi 
\im \frac{q}{p}}$ satisfies
 \begin{align}
  \text{order of pole}(\HS_{\G,F}(t))|_{t=e^{2\pi 
\im \frac{q}{p}}} &\leq \max_{\tau \in F}\left\{ 
\#\left\{ b_j(t) \, \big| \; b_j(\tau)\in p \cdot \NN \, \right\} \right\} 
+ \#\left\{ x \in X \, \big| \, \Delta(x)\in p \cdot \NN \right\} \notag \\
&\leq r + \#\left\{ x \in X \, \big| \, \Delta(x)\in p \cdot \NN \right\} \; ,
\end{align}
\end{myProp}
Proving this estimates starts from~\eqref{eqn:HS_pole_calc} wherein the 
contribution of each cone is now estimated by
\begin{align} 
Z_\tau\coloneqq \#\left\{ b_j(\tau) \, \big| \; b_j(\tau)\in p \cdot \NN \, 
\right\} 
&+ \#\left\{ x \in X_\tau \, \big| \, \Delta(x)\in p \cdot \NN \right\} \\
&- \text{order of zero} \left\{ \sum_{k=1}^{d_\mathrm{max}(\tau)}  
\mathrm{card}(B_k)  t^{k} \, \big| \, t=e^{2\pi 
\im \frac{q}{p}} \right\} \notag \; .
\end{align}
The individual pieces can be restrained as 
\begin{subequations}%
\begin{align}%
 0 &\leq \#\left\{ b_j(\tau) \, \big| \; b_j(\tau)\in p \cdot \NN \, 
\right\} \leq r \; , \\
0 &\leq \#\left\{ x \in X_\tau \, \big| \, \Delta(x)\in p \cdot \NN \right\} 
\leq \#\left\{ x \in X \, \big| \, \Delta(x)\in p \cdot \NN \right\} \; , \\
0&\leq \text{order of zero} \left\{ \sum_{k=1}^{d_\mathrm{max}(\tau)}  
\mathrm{card}(B_k)  t^{k} \, \big| \, t=e^{2\pi 
\im \frac{q}{p}} \right\} \leq \frac{d_\mathrm{max}(\tau) 
-1}{2} \; .
\end{align}%
\label{eqn:estimates_other_units}%
\end{subequations}%
The last estimate relies on the fact that non-real roots of unity have to 
appear as complex conjugated zeros of a real polynomial. 
Putting all the pieces together allows to derive at the desired claim.
% 
%%%%%%%%%%%%%%%%%%%%%%%%%%%%%%%%%%%%%%%%%%%%%%%%%%%%%%%%%%%%%%%%%%%%%%%%%%%%%%%%
  \section{Chiral ring generators}
\label{sec:chiral_ring}
The introduction of fans, monoids, and Hilbert bases has not only the 
computational benefits for the monopole formula as Hilbert series of the 
Coulomb branch $\MCoulomb$, it additionally allows us to provide a sufficient 
set of chiral ring generators.

\begin{myProp}
\label{prop:chiral_ring}
 The chiral ring $\C[\MCoulomb]$ for a theory with rank $r$ gauge group $\G$ 
and fan $F$ is generated by the following 
set of dressed monopole operators:
\begin{equation}
 \bigcup_{x \in \Hilb(F)} \left\{V_x \cdot g \, \big| \, g\in 
M_{x}^{\text{Dress}} \right\}  
\cup 
\left\{V_0 \cdot f_j \, \big| \, j=1,\ldots r \right\}
\; ,
\label{eqn:gen_chiral_ring}
\end{equation}
where $\Hilb(F) =  \cup_{\tau \in F} \Hilb(S_\tau)$ is the set of all monoid 
generators, and $\{f_1,\ldots,f_r\}$ are the Casimir 
invariants of $\G$. Moreover, $V_0$ denotes the trivial monopole 
operator with vanishing magnetic charge and zero 
conformal dimension.
\end{myProp}
To prove this claim, we proceed in steps. Firstly, notice that the 
(non-trivial) bare 
monopole operators are necessarily generated by $V_x$ for all $x \in \Hilb(F)$ 
by construction of the fan and the properties of the Hilbert bases. The bare 
monopole operators are, of course, a subset of~\eqref{eqn:gen_chiral_ring}, 
obtained by choosing $g=1$ in each $M_{x}^{\text{Dress}}$.

Secondly, the Casimir invariants $\{f_1,\ldots,f_r\}$ are the $r$ algebraic 
independent generators of $\mathfrak{J}(\hfrak)^{\Wcal_{\G}}$, which is the 
allowed polynomial algebra for the dressing of the origin in the magnetic 
weight lattice.

Thirdly, for a given bare monopole operator $V_x$ the algebra of dressing 
factors is equivalent to $\mathfrak{J}(\hfrak_m)^{\Wcal_{\Hh_m}}$. Moreover, 
the Casimir invariants of $\G$ span the ideal 
$\widetilde{I}_{\G}^{\mathrm{max}} \subset 
\mathfrak{J}(\hfrak_m)^{\Wcal_{\Hh_m}}$. Therefore, the only elements in $\{ 
V_x \cdot h\,  |\;  h \in \mathfrak{J}(\hfrak_m)^{\Wcal_{\Hh_m}} \}$ which are 
not generated by $V_x \cdot f $ for $f \in \widetilde{I}_{\G}^{\mathrm{max}} $ 
are precisely of the form $V_x \cdot g $ for $g \in M_{x}^{\text{Dress}} $. 
Hence, the proposition holds.

As a remark, this set of generators is really to be understood as sufficient, 
but in most cases it will not be necessary. That is, we have not yet identified 
the relations between these operators, which is still an open problem.

\begin{figure}[h]
\centering
\begin{tikzpicture}
  \coordinate (Origin)   at (0,0);
  \coordinate (XAxisMin) at (0,0);
  \coordinate (XAxisMax) at (5,0);
  \coordinate (YAxisMin) at (0,0);
  \coordinate (YAxisMax) at (1.2*5,1.2*2.5);
  \draw [thin, gray] (XAxisMin) -- (XAxisMax);%
  \draw [thin, gray] (YAxisMin) -- (YAxisMax);%
  \draw (-0.1,-0.25) node {$x_0$};
  \draw (1.15,0.76) node {$x_1$};
  \draw (3.0,-0.25) node {$x_2$};
  \draw (4.66,1.5*0.66-0.25) node {$x_3$};
  \draw (0.78*5.2,0.78*2.6+0.25) node {$S_1$};
  \draw (4.6,-0.25) node {$S_2$};
  \draw (5.6,1.25) node {$S$};
%Draw the monoid
  \foreach \x in {0}{%
        \node[draw,circle,inner sep=0.8pt,fill,black] at (\x,0) {};
            }
    \foreach \x in {1,1.25,...,1.5}{%
        \node[draw,circle,inner sep=0.8pt,fill,black] at (\x,1.5*\x-1.5) {};
        \node[draw,circle,inner sep=3pt,thick,black] at (\x,1.5*\x-1.5) {};
            }            
    \foreach \x in {1,1.33,...,1.99}{%
        \node[draw,circle,inner sep=0.8pt,fill,black] at (1+\x,1.5*\x-1.5) {};
            }            
    \foreach \x in {1,1.5,...,2.5}{%
        \node[draw,circle,inner sep=0.8pt,fill,black] at (2+\x,1.5*\x-1.5) {};
            }            
    \foreach \x in {1,1.66,...,2.99}{%
        \node[draw,circle,inner sep=0.8pt,fill,black] at (3+\x,1.5*\x-1.5) {};
            }        
%             
%draw the Casimir invariance over the origin
  \draw [thin,dashed, gray] (Origin) -- (0,6.2);%
  \foreach \x in {3,5,6}{%
        \node[draw,diamond,inner sep=0.8pt,fill,red] at (0,\x) {};
            }
%             
%draw Casimir invariance over first boundary
  \draw [thin,dashed, gray] (1.5,0.75) -- (1.5,0.75+6.2);%
  \foreach \x in {3,5,6}{%
        \node[draw,diamond,inner sep=0.8pt,fill,red] at (1.5,0.75+\x) {};
        }
  \foreach \x in {2,4}{%
        \node[draw,cross out,thick,inner sep=1pt,blue] at 
(1.5,0.75+\x) {};
        }
%         
%draw Casimir invariance over second boundary
  \draw [thin,dashed, gray] (3,0) -- (3,6.2);%
  \foreach \x in {3,5,6}{%
        \node[draw,diamond,inner sep=0.8pt,fill,red] at (3,\x) {};
        }
  \foreach \x in {4}{%
        \node[draw,cross out, thick,inner sep=1pt,blue] at 
(3,\x) {};
        }
%         
%draw Casimir invariance over interior
  \draw [thin,dashed, gray] (4.66,1.5*0.66) -- (4.66,6.2+1.6*0.66);%
  \foreach \x in {3,5,6}{%
        \node[draw,diamond,inner sep=0.8pt,fill,red] at (4.66,\x+1.5*0.66) {};
        }
  \foreach \x in {1,2,4}{%
        \node[draw,cross out, thick,inner sep=1pt,blue] at (4.66,\x+1.5*0.66) 
{};
        }
\end{tikzpicture}%
\caption{A rank $2$ monoid $S$ with Hilbert basis given by the circled points. 
The ``tower'' over the four representative points $x_0,\ldots, x_3$ indicates 
the dressed monopole operators. The red points denote the dressing by Casimir 
invariants of the gauge group $\G$, while the blue crosses represent the 
(finite number of) dressings by the residual gauge group $\Hh_x$.}%
\label{fig:chiral_ring}%
\end{figure}
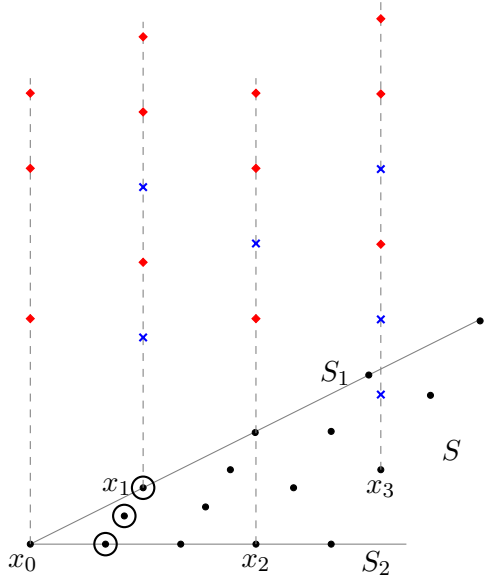

Lastly, let us illustrate Prop.~\ref{prop:chiral_ring} in 
Fig.~\ref{fig:chiral_ring}.
Displayed is a monoid $S$ of rank $2$, which can be decomposed as
\begin{equation}
 S = \{x_0\} \uplus \Relint(S_1) \uplus \Relint(S_2) \uplus \Relint(S) \; .
\end{equation}
The four points originate from different parts: $x_1\in \Relint(S_1)$, $x_2 \in 
\Relint(S_2)$ and $x_3 \in \Relint(S)$. The pile of red diamonds over $x_0$ 
denotes the dressed monopole operators of the trivial operator $V_{x_0}$, i.e. 
the Casimir invariants $\{f_1,f_2\}$ of $\G$ itself. In general, the degrees 
satisfy 
$\deg(f_j)>1$, such that not all (small) degrees are realised in 
$\C[f_1,f_2]$. 
The bare monopole operator $V_{x_1}$ can be dressed by the Casimir invariants 
of $\G$, which is represented by the pile of red diamonds over $x_1$. However, 
the residual gauge group is a subgroup of $\G$, such that Casimir 
invariance is generated by $\{g_1,g_2\}$ with different degrees as the $f_j$. 
The elements in $\C[g_1,g_2]$ which are not simultaneously in 
$\widetilde{I}_{\G}^{\mathrm{max}}$ have to lie in the quotient 
$M_{x_1}^{\text{Dress}}$. These contributions correspond to the blue crosses in 
the ``tower'' over $x_1$. Since the Hilbert series of $M_{x_1}^{\text{Dress}}$ 
is finite, this means that only a finite number of dressed monopole operators 
associated to $V_{x_1}$ are not expressed as product of $V_{x_1}$ and Casimir 
invariants of $\G$. 

The very similar behaviour occurs for $x_2$ and $x_3$. Note however, that the 
residual gauge groups can vary even between monoids with the same rank, here 
$S_1$ and $S_2$. Moreover, the relative interior of $S$, which here is assumed 
to be a subset of the relative interior of the dominant Weyl chamber, exhibits 
the largest ``tower'' of dressed operators, as the degrees of the Casimir 
invariants are $1$.
% 
%%%%%%%%%%%%%%%%%%%%%%%%%%%%%%%%%%%%%%%%%%%%%%%%%%%%%%%%%%%%%%%%%%%%%%%%%%%%%%%%
  \section{Implementation}
\label{sec:implementation}
In this section we show how the concepts described in the earlier sections 
allow for an implementation in software like \texttt{Macaulay2} and 
\texttt{Mathematica}. Thus, we show how to apply the methods to more physical 
theories, such as certain quiver gauge theories.
\subsection{A recipe}
Given a $3$-dimensional $\Ncal=4$ gauge theory with gauge group $\G$ and some 
matter content, there are five steps for computing the Hilbert series with a 
combined use of \texttt{Macaulay2} and \texttt{Mathematica}. (There may be 
other programs capable of performing the same tasks. Here, we focus on those 
two.)
\begin{enumerate}[(I)]
 \item \textbf{Definition of the fan:} One approach is based on the 
\texttt{Macaulay2} package \emph{Polyhedra}~\cite{Polyhedra:Birkner}, through 
which cones can be defined 
by intersection of hyper-planes. This is precisely the approach we have advocated 
in~\cite{Hanany:2016ezz} and elaborated on in the previous sections.

Therefore, one identifies all relevant weights in the conformal dimension of 
the theory and computes the hyper-planes which define the dominant Weyl chamber 
of the GNO-dual group. Next, the intersection of suitable subsets defines all 
cones, which can be grouped together as a fan $F$.
\item \textbf{Computation of the toric ideal:} For each cone $\tau\in F$ one 
computes the Hilbert basis $\Hilb(S_\tau)$ with respect to the lattice 
$\Lambda\cong \Z^{\rank(G)}$ via \texttt{Macaulay2}.

The output $\Hilb(S_\tau)$ is graded via $\Gamma_{\Delta}$ and the resulting 
monoid generators are fed into the algorithm described in~\emph{Algorithm for 
the Toric Hilbert Scheme} by M.\ Stillman, B.\ Sturmfels, and R.\ Thomas 
in~\cite[Part II]{Macaulay2book}. The output is the lattice (or toric) ideal 
for the corresponding monoid.

The Hilbert series for the lattice ideal as well as the quotient ring can be 
straight forwardly evaluated in~\texttt{Macaulay2}.
\item \textbf{Casimir invariance:} The \Poincare\ series 
$\PoinS_{\Relint(S_\tau)}(t)$ can be evaluated as Hilbert series of the free 
polynomial ring $\C[f_1,\ldots,f_r]$ wherein one only has to define the degrees 
$\deg(f_j)=b_j$ according to the residual gauge group $\Hh_m$.
\item \textbf{Combinatorics:} With all the aforementioned ingredients, the 
final step consists of summing over all cones in the fan $F$, while for each 
cone $\tau$ we need to employ the exclusion-inclusion principle as shown 
in~\eqref{eqn:exclusion_inclusion}. Therefore, the contribution of each cone 
$\tau$ involves a sum over $\Fcal(\tau)$.
\item \textbf{Algebraic manipulations:} The \texttt{Macaulay2} output can be 
exported into a \texttt{Mathematica} 
readable format, which allows for the algebraic manipulations such as taking 
the limit~\eqref{eqn:HS_def_limit}, bringing the sum into the form of an 
rational function, or analysing the result with the plethystic logarithm.
\end{enumerate}
% 
%%%%%%%%%%%%%%%%%%%%%%%%%%%%%%%%%%%%%%%%%%%%%%%%%%%%%%%%%%%%%%%
%%%%%%%%%%%%%%%%%%%%%%%%%%%%%%%%%%%%%%%%%%%%%%%%%%%%%%%%%%%%%%%
%
\subsection{Example quiver gauge theories}
Now, we demonstrate how our approach applies to some exemplary quiver gauge 
theories. We highlight in each example the geometric content of the fan and 
show that the sum over all cones can become cumbersome. However, the benefit is 
that the fan provides a structuring pattern which allows a well-defined and 
concise approach to the monopole formula in contrast to the pure brute force 
evaluation without this information.
\subsubsection{Example I: rank 3}
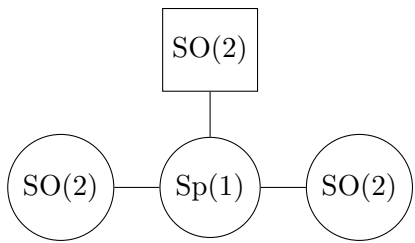
\begin{figure}[h]
\centering
\begin{tikzpicture}%[font=\footnotesize]
\begin{scope}[auto,%
  every node/.style={draw, minimum size=1.1cm}, node distance=0.6cm];
  % the vertices
\node[circle] (SO1) at (0, 0) {$\sorm(2)$};
\node[circle, right=of SO1] (C1) {$\sprm(1)$};
\node[circle, right=of C1] (SO2) {$\sorm(2)$};
\node[rectangle, above=of C1] (SOF) {$\sorm(2)$};
\end{scope}
  % the edges
\draw (SO1) -- (C1)
(C1)--(SO2)
(C1)--(SOF);
\end{tikzpicture}
\caption{The quiver diagram for the first example. The magnetic charges for the 
two $\sorm(2)$ factors are labelled by $n_i\in \Z$ for $i=1,2$, while the 
$\sprm(1)$ magnetic weights are denoted by $m\in \NN$.}
\label{fig:quiver_SO2-C1-SO2}
\end{figure}

We study the quiver gauge theory displayed in Fig.~\ref{fig:quiver_SO2-C1-SO2}.
By the well-known fact $C_1\cong A_1$, we can treat $\sprm(1)$ effectively 
as $\su$. 
The magnetic weights of the gauge group are parametrised by $n_j 
\in \Z$, $j=1,2$, for each of the two $\sorm(2)$ factors, and $m \in 
\NN$ for the $\sprm(1)\cong\su$ factor. It readily follows that the 
corresponding dominant Weyl chamber $\sigma$ of the GNO-dual group is an 
entire half-space in $\R^3$, i.e.\ $\sigma$ is the $\R_+$ span of the rank $3$ 
monoid $\{(n_1,m,n_2) \in \Z^3 | n_j \geq 0 \, , \, j=1,2\}$. Therefore, 
$\sigma$ is not strongly convex cone, which is not surprising as the gauge 
group is not semi-simple because $\sorm(2)\cong \uo$.

The conformal dimension of the quiver gauge theory reads
 \begin{equation}
  \Delta_{\text{Ex.I}}(n_1,m,n_2)= \frac{1}{2} \sum_{j=1}^{2} \left( |n_j+m| + 
|n_j-m|\right) 
 +|m| - 2|m| \; .
 \end{equation}
Employing this information, one defines the fan $F_{\text{Ex.I}}$ inside 
$\sigma$ via the following $4$ hyper-planes in $\R^3$:
\begin{equation}
 n_j \pm m =0 \; , \qquad \for j =1,2 \; .
\end{equation}
It turns out that the matter content is enough to generate a fan 
consisting of positive rational cones only. The number and dimensions of the 
cones in $F_{\text{Ex.I}}$ are provided in Tab.~\ref{tab:dim_cones_SO2-C1-SO2}.
\begin{table}[h]
\centering
\begin{tabular}{c|c|c|c|c}
\toprule
  $\dim(\tau)$& 0 & 1 & 2 & 3  \\ \midrule
   $\#(\tau)$, $\tau\in F_{\text{Ex.I}}$ & 1 & 8 & 16 & 9 \\
   \bottomrule
\end{tabular}
\caption{Number of cones in the $F_{\text{Ex.I}}$ for given dimension.}
\label{tab:dim_cones_SO2-C1-SO2}
\end{table}
The dressing factors for the gauge group $\sorm(2)\times \sprm(1)\times 
\sorm(2)$ are easily accounted for, because only the $\sprm(1)$ 
contribution is non-constant in $\sigma$. Thus, one obtains
\begin{equation}
 P_{\text{Ex.I}}(t^2;n_1,m,n_2) =P_{\sprm(1)}(t^2;m) \cdot \prod_{i=1}^2 
P_{\sorm(2)} (t^2;n_i) =  \begin{cases} 
 \frac{1}{1-t^4} \cdot \frac{1}{(1-t^2)^2} & m=0 \; , \\
 \frac{1}{1-t^2} \cdot \frac{1}{(1-t^2)^2} & m>0 \; .\\                     
                    \end{cases}
\end{equation}
The collection of all Hilbert bases elements for the fan $F_{\text{Ex.I}}$ is 
displayed 
Tab.~\ref{tab:minimal_generators_SO2-C1-SO2}. These minimal monoid generators 
correspond to the bare monopole operators of the 
theory in Fig.~\ref{fig:quiver_SO2-C1-SO2}.
\begin{table}[h]
 \centering
 \begin{tabular}{c|c|c}
 \toprule
  \multicolumn{2}{c|}{$(n_1,m,n_2,2\Delta)$} & 
$\PoinS_{M_{(n_1,m,n_2)}^{\text{Dress}}}(t^2) $ \\ 
\midrule
$(1,0,0,2)$ & $(-1,0,0,2)$ & $1$ \\
$(0,0,1,2)$ & $(0,0,-1,2)$ & \\ \midrule
$(0,1,0,2)$ & $(1,1,0,2)$ & $1+t^2$ \\
$(-1,1,0,2)$ & $(0,1,1,2)$ & \\
$(0,1,-1,2)$ & $(1,1,1,2)$ & \\
$(-1,1,1,2)$ & $(1,1,-1,2)$  & \\
$(-1,1,-1,2)$ & & \\ \midrule
\multicolumn{3}{c}{+ $2$ Casimir invariants of degree $2$ } \\ \midrule
\multicolumn{3}{c}{+ $1$ Casimir invariant of degree $4$} \\
\bottomrule
 \end{tabular}
 \caption{The minimal generators plus their dressing behaviour of the 
quiver gauge theory displayed in Fig.~\ref{fig:quiver_SO2-C1-SO2}.}
\label{tab:minimal_generators_SO2-C1-SO2}
\end{table}
After some algebraic manipulations the Hilbert series, using additional grading 
$z_i^{n_i}$ for the two 
$\sorm(2)$ factors, reads as follows:
\begin{subequations}
\begin{align}
\HS_{\text{Ex.I}}(t^2,z_1,z_2)=& \frac{R_{\text{Ex.I}}(t^2,z_1,z_2) }{
P_{\text{Ex.I}}(t^2,z_1,z_2) } \\
R_{\text{Ex.I}}(t^2,z_1,z_2)=& 1 
+ \left(3+ \frac{1}{z_1} +z_1 + \frac{1}{z_2} +z_2\right) t^2 \\
&\phantom{1} 
-\left( \frac{1}{z_1} +z_1 + \frac{1}{z_2} +z_2 + \frac{1}{z_1 z_2} + z_1 
z_2 + \frac{z_1}{z_2} + \frac{z_2}{z_1}\right) t^4 \notag\\
&\phantom{1} 
-\left( \frac{1}{z_1} +z_1 + \frac{1}{z_2} +z_2 +\frac{1}{z_1 z_2} + z_1 z_2 + 
\frac{z_1}{z_2} + \frac{z_2}{z_1} \right) t^6 \notag\\
&\phantom{1} 
+ \left(3+ \frac{1}{z_1} +z_1 + \frac{1}{z_2} +z_2\right) t^8 + t^{10} \notag \\
P_{\text{Ex.I}}(t^2,z_1,z_2) =&
\left(1 - \frac{1}{z_1} t^2 \right) 
\left(1 - z_1 t^2  \right) 
\left(1 - \frac{1}{z_2} t^2 \right) 
\left(1 - z_2 t^2 \right) 
\left(1 - \frac{1}{z_1 z_2}t^2 \right) 
\left(1 - z_1 z_2 t^2 \right) \\*
&\phantom{\left(1 - \frac{1}{z_1} t^2 \right) \left(1 - z_1 t^2  \right) } 
\times \left(1 - \frac{z_1}{z_2} t^2 \right) \left(1 - \frac{z_2}{z_1} t^2 
\right) \notag \\
\HS_{\text{Ex.I}}(t^2) =& \lim_{z_1,z_2\to1} 
\HS_{\text{Ex.I}}(t^2,z_1,z_2) =\frac{1+9 t^2 +9 t^4+ t^6}{\left(1-t^2\right)^6}
 \label{eqn:HS_SO2-C1-SO2}
\end{align}
\end{subequations}
and subsequent expansion of the Hilbert series and the plythestic logarithm 
yields
\begin{subequations}
 \begin{align}
  \HS_{\text{Ex.I}}(t^2)&= 1 + \left(3 + \frac{2}{z_1} + 2 z_1 + \frac{2}{z_2} + 
2 z_2 + 
\frac{1}{z_1 z_2} +  z_1 z_2 + \frac{z_1}{z_2}  + \frac{z_2}{z_1} \right) t^2 
+ \mathcal{O}(t^{4})  \label{eqn:HS_SO2-C1-SO2_expanded}\\
&\xrightarrow{z_1,z_2\to1} 1 + 15 t^2 + 84 t^4 + 300 t^6 + 825 t^8 + 1911 
t^{10} + \mathcal{O}(t^{12}) \; ,\notag \\
  \PL_{\text{Ex.I}}(t^2)&=\left(3 + \frac{2}{z_1} + 2 z_1 + \frac{2}{z_2} + 2 
z_2 + 
\frac{1}{z_1 z_2} +  z_1 z_2 + \frac{z_1}{z_2}  + \frac{z_2}{z_1} \right) t^2
- \mathcal{O}(t^{4}) \label{eqn:PL_SO2-C1-SO2} \\
&\xrightarrow{z_1,z_2\to2} 15t^2 - 36 t^4 + 160 t^6 - 945 t^8 + 6048 
t^{10} - \mathcal{O}(t^{12}) \; , \notag
 \end{align}
\end{subequations}
We readily observe the following:
\begin{enumerate}[(i)]
 \item The Hilbert series~\eqref{eqn:HS_SO2-C1-SO2} has a pole of order $6$ as 
$t$, implying that the complex dimension of the moduli space is $6$.
  \item The difference in degrees of the denominator and the numerator 
of~\eqref{eqn:HS_SO2-C1-SO2} is $6$, which agrees with complex 
dimension of the Coulomb branch.
  \item The numerator is a palindromic polynomial.
  \item The expansion~\eqref{eqn:HS_SO2-C1-SO2_expanded} shows that the global 
symmetry group of the moduli space is of dimension $15$ and rank $3$. Two 
prominent possibilities are $\surm(4)$ and 
$\sorm(6)$, which are accidentally related via $\surm(4) \cong 
\mathrm{Spin}(6)$ and  $\sorm(6) \cong \mathrm{Spin}(6)\slash \Z_2$. This idea 
is supported by the following:
\begin{subequations}
\begin{align}
 \boldsymbol{15}_{\surm(4)} \big|_{\surm(3)} &= \boldsymbol{8}_{\surm(3)}
 + \boldsymbol{3}_{\surm(3)} + \boldsymbol{\bar{3}}_{\surm(3)} +  
\boldsymbol{1}_{\surm(3)}  \; ,
 \\
\HS\big|_{t^2}= \PL\big|_{t^2} 
% &= 3 + \frac{2}{z_1} + 2 z_1 + \frac{2}{z_2} + 2 
% z_2 + 
% \frac{1}{z_1 z_2} +  z_1 z_2 + \frac{z_1}{z_2}  + \frac{z_2}{z_1} \\
% 
&= \left[ 2 + \frac{1}{z_1} +  z_1 + \frac{1}{z_2} + 1 z_2 + 
\frac{1}{z_1 z_2} +  z_1 z_2 \right] \\
&\qquad + \left[ z_1 +\frac{z_2}{z_1} +\frac{1}{z_2 } \right] 
 + \left[ z_2 +\frac{z_1}{z_2} +\frac{1}{z_1 } \right] +1 \notag \\
&= \chi^{\surm(3)}(\boldsymbol{8}) +\chi^{\surm(3)}(\boldsymbol{3}) 
+\chi^{\surm(3)}(\boldsymbol{\bar{3}}) + \chi^{\surm(3)}(\boldsymbol{1}) \; .
\end{align}
\end{subequations}
  \item The PL~\eqref{eqn:PL_SO2-C1-SO2} reveals $15$ generators at degree $2$. 
These are given by the $13$ minimal generators 
of Tab.~\ref{tab:minimal_generators_SO2-C1-SO2} plus the two degree $2$ Casimir 
invariants of the gauge group. Note that the magnetic weights agree with the 
$z_1$, $z_2$-grading.
  \item We see neither the degree $4$ Casimir invariant nor the $9$ dressed 
monopole operators of Tab.~\ref{tab:minimal_generators_SO2-C1-SO2} in the PL. 
However, it does not necessarily imply that these generators are not present.
\end{enumerate}
Lastly, we note that the moduli space of the theory encoded in 
Fig.~\ref{fig:quiver_SO2-C1-SO2} is the closure of the minimal nilpotent orbit 
of $\surm(4)$ and, hence, can be compared to the study of the reduced one $\surm(4)$ 
instanton moduli space of~\cite{Benvenuti:2010pq}.
%
%%%%%%%%%%%%%%%%%%%%%%%%%%%%%%%%%%%%%%%%%%%%%%%%%%%%%%%%%%%%%%%%%%%%%%%%
%%%%%%%%%%%%%%%%%%%%%%%%%%%%%%%%%%%%%%%%%%%%%%%%%%%%%%%%%%%%%%%%%%%%%%%%
%
\subsubsection{Example II: rank 4}
\begin{figure}[h]
\centering
\begin{tikzpicture}%[font=\footnotesize]
\begin{scope}[auto,%
  every node/.style={draw, minimum size=1.1cm}, node distance=0.6cm];
  % the vertices
\node[circle] (SO2) at (0, 0) {$\sorm(2)$};
\node[circle, right=of SO2] (C1) {$\sprm(1)$};
\node[circle, right=of C1] (SO4) {$\sorm(4)$};
\node[rectangle, above=of SO4] (SpF) {$\sprm(2)$};
\end{scope}
  % the edges
\draw (SO2) -- (C1)
(C1)--(SO4)
(SO4)--(SpF);
\end{tikzpicture}
\caption{The quiver diagram for the second example. The magnetic charge for 
$\sorm(2)$ is labelled by $n\in \Z$ and for $\sprm(1)$ by $m\in \NN$. The 
two magnetic weights of $\sorm(4)$ are denoted by $p_1,p_2\in \Z$, which 
satisfy $p_1\geq |p_2|\geq 0$. }
\label{fig:quiver_SO2-C1-SO4}
\end{figure}
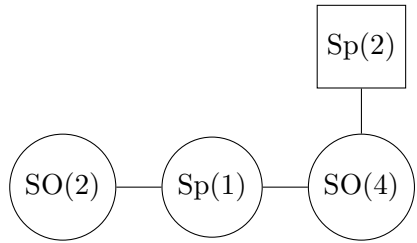

Let us now examine the quiver gauge theory of Fig.~\ref{fig:quiver_SO2-C1-SO4}. 
The monoid, resulting from the intersection of the dominant Weyl chamber of the 
GNO-dual group and the magnetic weight lattice, is parametrised via $n\in \Z$, 
$m \in \NN$, and $p_1,p_2 \in \Z$ with $p_1\geq |p_2|\geq0$. This rank $4$ 
monoid is non-positive due to the $\sorm(2)\cong \uo$ factor in the gauge 
group. Hence, the Weyl chamber as rational cone thereof is non-positive cone in 
$\R^4$.

The conformal dimension of the theory under consideration reads as
 \begin{equation}
 \begin{aligned}
  \Delta_\mathrm{Ex.II}(n,m,p_1,p_2)= \frac{1}{2} \left( |n-m| + |n+m|\right) &+
  \frac{1}{2} \sum_{j=1}^{2} 
\left( |m+p_j| + |m-p_j|\right) 
 +2\left(|p_1| + |p_2| \right) \\
 &-2|m| - |p_1+p_2| - |p_1-p_2| \; .
\end{aligned}
 \end{equation}
As before, we deduce the fan $F_\mathrm{Ex.II}$ via the following hyper-planes 
in $\R^4$:
\begin{equation}
 n\pm m=0 \; , \quad  m+p_2=0 \; , \quad  p_2=0 \; , \und m-p_j=0 \for j=1,2 \; 
.
\end{equation}
From hyper-plane arrangement we compute the fan, whose cones are 
summarised in Tab.~\ref{tab:dim_cones_SO2-C1-SO4}.
\begin{table}[h]
\centering
\begin{tabular}{c|c|c|c|c|c}
\toprule
  $\dim(\tau)$& 0 & 1 & 2 & 3 & 4 \\ \midrule
   $\#(\tau)$, $\tau\in F_\mathrm{Ex.II}$ & 1 & 13 & 40 & 46 & 18 \\
   \bottomrule
\end{tabular}
\caption{Number of cones in the fan $F_\mathrm{Ex.II}$ for given dimension.}
\label{tab:dim_cones_SO2-C1-SO4}
\end{table}
Again, the matter content is enough to generate a fan that consists solely of 
strongly convex cones such that all monoids are positive. 
Because of the positivity of all monoids, we can compute the 
Hilbert bases and the corresponding dressings. The results are displayed in 
Tab.~\ref{tab:minimal_generators_SO2-C1-SO4}.
\begin{table}[h]
\centering
\begin{tabular}{c|c|c}
 \toprule
\multicolumn{2}{c|}{$(n,m,p_1,p_2,2\Delta)$} & 
$\PoinS_{M_{(n,m,p_1,p_2)}^{\text{Dress}}}(t^2)$ \\ 
\midrule 
%degree 2
$(1,0,0,0,2)$ & $(-1,0,0,0,2)$ & $1$ \\ \midrule
$(0,1,0,0,2)$ & $(1,1,0,0,2)$ & $1+t^2$ \\
$(-1,1,0,0,2)$  & & \\ \midrule
$(0,0,1,0,2)$ & &  $1+2 t^2+t^4$ \\ \midrule
$(0,1,1,0,2)$ & $(1,1,1,0,2)$ & $1+3 t^2+3 t^4+t^6$ \\
$(-1,1,1,0,2)$ & & \\ \midrule
\multicolumn{3}{c}{+ $1$ Casimir invariant of degree $2$} \\ \midrule
%degree 4
\multicolumn{3}{c}{+ $3$ Casimir invariants of degree $4$} \\ \midrule
%degree 6
$(0,1,1,1,6)$ & $(0,1,1,-1,6)$ & $1+2 t^2+t^4$ \\
$(1,1,1,1,6)$ & $(-1,1,1,1,6)$ & \\
$(1,1,1,-1,6)$ & $(-1,1,1,-1,6)$ & \\ \midrule
%degree 8
$(0,0,1,1,8)$ & $(0,0,1,-1,8)$ & $1+t^2$ \\
\bottomrule
\end{tabular}
\caption{The set of minimal generators plus their dressing behaviour of the 
quiver gauge theory displayed in Fig.~\ref{fig:quiver_SO2-C1-SO4}.}
\label{tab:minimal_generators_SO2-C1-SO4}
\end{table}
In addition, the relevant classical dressing factors, which are the products of 
the 
dressing factors for each factor in the quiver gauge group, are given by
\begin{equation}
 P_{\text{Ex.II}}(t^2;n,m,p_1,p_2)= 
 P_{\sorm(2)}(t^2;n) \cdot P_{\sprm(1)}(t^2;m) \cdot 
P_{\sorm(4)}(t^2;p_1,p_2) \; .
\end{equation}
The dressing factors for the special orthogonal groups and for $\sprm(1)\cong 
\su$ are provided in~\cite{Cremonesi:2013lqa}.
The resulting Hilbert series, wherein we choose to additionally grade the 
$\sorm(2)\cong\uo$ factor by $z$, is evaluated to
\begin{equation}
 \HS_{\text{Ex.II}}(t^2,z)=\frac{1 + 2 t^2 + 2 t^4 +  2 t^6 + t^8
 }{\left(1 - t^2\right)^2 \left(1 - \frac{1}{z} t^2\right)^3 \left(1 - z t^2 
\right)^3} 
= \frac{\left(1 - t^4\right) \left(1 - t^8\right)}{\left(1 - t^2\right)^4 
\left(1 - \frac{1}{z} t^2\right)^3 \left(1 - z t^2 \right)^3} 
\label{eqn:HS_SO2-C1-SO4}
 \end{equation}
and expanding the Hilbert series and plythestic logarithm yields
\begin{subequations}
 \begin{align}
  \HS_{\text{Ex.II}}(t^2,z)=& 1 + \left[4 + 3\left(z+\tfrac{1}{z}\right) \right] 
t^2 
  + \left[18 + 12\left(z+\tfrac{1}{z}\right) + 6\left(z^2+\tfrac{1}{z^2}\right) 
  \right] t^4 \label{eqn:HS_SO2-C1-SO4_expand} \\
  &+ \left[52+ 45 \left(z+\tfrac{1}{z}\right) + 24 
\left(z^2+\tfrac{1}{z^2}\right) +     10 \left(z^3+\tfrac{1}{z^3}\right)\right] 
t^6 
+ \mathcal{O}(t^{8}) \notag \; , \\
 \PL_{\text{Ex.II}}(t^2,z) =& \left(4 + 3\tfrac{1}{z} + 3 z\right) t^2 - t^4 - 
t^8\; . 
\label{eqn:PL_SO2-C1-SO4}
 \end{align}
\end{subequations}
We find the following:
\begin{enumerate}[(i)]
 \item The Hilbert series~\eqref{eqn:HS_SO2-C1-SO4} has a pole of order $8$ at 
$t\to 1$, 
i.e.\ the moduli space is of complex dimension $8$. 
  \item The difference in degrees between denominator and numerator is $8$, 
agreeing with the dimension of the Coulomb branch as well.
  \item The PL~\eqref{eqn:PL_SO2-C1-SO4} shows ten generators at degree $2$, 
these agree with the nine minimal generators plus one Casimir invariant of 
Tab.~\ref{tab:minimal_generators_SO2-C1-SO4}. Note also that the $z$-grading 
agrees with the magnetic charges of the minimal generators.
  \item The PL~\eqref{eqn:PL_SO2-C1-SO4} displays a relation at degree $4$ 
as well as one relation as degree $8$. This agrees with the complete 
intersection form of~\eqref{eqn:HS_SO2-C1-SO4}.
\end{enumerate}
Finally, let us note that the moduli space of the quiver gauge theory encoded 
in Fig.~\ref{fig:quiver_SO2-C1-SO4} is the closure of the maximal nilpotent 
orbit of $\sorm(5)$.
%
%%%%%%%%%%%%%%%%%%%%%%%%%%%%%%%%%%%%%%%%%%%%%%%%%%%%%%%%%%%%%%%%%%%%%%
%%%%%%%%%%%%%%%%%%%%%%%%%%%%%%%%%%%%%%%%%%%%%%%%%%%%%%%%%%%%%%%%%%%%%%
%
\subsubsection{Example III: rank 6}
\begin{figure}[h]
\centering
\begin{tikzpicture}%[font=\footnotesize]
\begin{scope}[auto,%
  every node/.style={draw, minimum size=1.1cm}, node distance=0.6cm];
  % the vertices
\node[circle] (O2L) at (0, 0) {$\orm(2)$};
\node[circle, right=of O2L] (C1L) {$\sprm(1)$};
\node[circle, right=of C1L] (O4) {$\orm(4)$};
\node[circle, right=of O4] (C1R) {$\sprm(1)$};
\node[circle, right=of C1R] (O2R) {$\orm(2)$};
\node[rectangle, above=of O4] (CF) {$\sprm(1)$};
\end{scope}
  % the edges
\draw (O2L) -- (C1L)
(C1L) -- (O4)
(O4) -- (C1R)
(C1R) -- (O2R)
(O4)--(CF);
\end{tikzpicture}
\caption{The quiver diagram for the third example. The magnetic charges for 
the $\orm(2)$-factors are labelled by $n_1,n_2\in \Z$ and for the 
$\sprm(1)$-factors by $m_1,m_2\in \NN$. The two magnetic weights of $\orm(4)$ 
are denoted by $p_1,p_2\in \NN$, which satisfy $p_1\geq p_2\geq 0$. }
\label{fig:quiver_O2-C1-O4-C1-O2}
\end{figure}
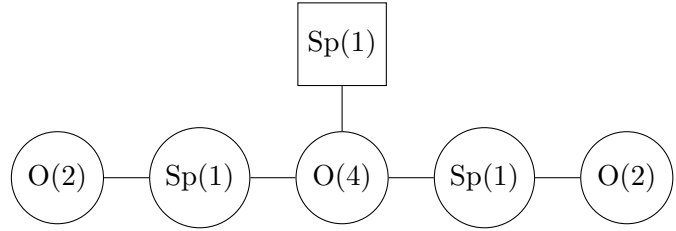

For the last example we choose a rank $6$ quiver gauge theory as displayed in 
Fig.~\ref{fig:quiver_O2-C1-O4-C1-O2}.
Let us start by commenting on some subtleties of the gauge group and its 
associated magnetic weight lattice. For the orthogonal groups  we proceed as 
follows: The  fundamental representation of $\orm(2)$ stems from 
$\sorm(2)$. Moreover, $\orm(2)$ is an abelian group and as such has no 
contributions from vector multiplets. In contrast, the GNO-dual and the dressing 
factors are those of $\sorm(3)$, as known from~\cite{Cremonesi:2014uva}.
Next, $\orm(4)$ behaves similarly, as the fundamental representation is 
inherited from $\sorm(4)$, but the GNO magnetic weight lattice and the dressing 
factors are those of $\sorm(5)$, by results of~\cite{Cremonesi:2014uva}.
Then, the monoid describing the magnetic weights in the dominant Weyl chamber is 
characterised by $n_1,n_2\in \Z$, $m_1,m_2\in \NN$, and $p_1,p_2\in \NN$ with 
$p_1\geq p_2 \geq0$. 

The conformal dimension of this gauge theory reads
\begin{equation}
 \begin{aligned}
  \Delta_{\text{Ex.III}}(n_1,m_1,p_1,p_2,m_2,n_2)=& 
  \frac{1}{2} \sum_{i=1}^2 \left( |m_i +n_i| + |m_i -n_i| \right) \\
  &+ \frac{1}{2}  \sum_{i,j=1}^2 \left(|m_i +p_j| + |m_i -p_j| \right) 
+\sum_{j=1}^2 |p_j| \\
&- 2\sum_{i=1}^2 |m_i| - |p_1+p_2| - |p_1-p_2| \; .
 \end{aligned}
 \label{eqn:ConfDim_O-type}
\end{equation}
We identify $5$ hyper-planes in $\R^6$ intersecting the GNO Weyl chamber 
$\sigma$ non-trivially. These are defined via
\begin{equation}
 m_i -n_i =0 \und m_i-p_j =0 \for i=1,2 \, , \;  j=1,2\; .
\end{equation}
Employing \texttt{Macaulay2}, one obtains a variety of cones that can be 
arranged into a fan $F_{\text{Ex.III}}$ and we summarise the cones in 
Tab.~\ref{tab:dim_cones_O-type}.
\begin{table}[h]
\centering
\begin{tabular}{c|c|c|c|c|c|c|c}
  $\dim(\tau) $& 0 & 1 & 2 & 3 & 4 & 5 & 6 \\ \hline
    $\#(\tau)$, $\tau \in F_{\text{Ex.III}}$ & 1 & 24 & 122 & 268 
& 297 & 164 & 36
\end{tabular}
\caption{Number of cones in the fan $F_{\text{Ex.III}}$ for given 
dimension.}
\label{tab:dim_cones_O-type}
\end{table}
Similar to the previous examples, the dominant Weyl chamber of the gauge group 
$\orm(2)^2 \times \sprm(1)^2 \times \orm(4)$ is not strongly convex. However, 
the fan generated by the matter content consists only of strongly convex cones 
such that the notion of Hilbert basis is applicable to the resulting positive 
monoids. Hence,  we computed the Hilbert bases $\Hilb(\tau)$ 
for every cone $\tau \in F_{\text{Ex.III}}$. 
The union of the sets is summarised in Tab.~\ref{tab:min_generators_quiver}.
\begin{table}[h]
\centering
\begin{tabular}{c|c|c}
\toprule
\multicolumn{2}{c|}{$(n_1,m_1,p_1,p_2,m_2,n_2,2\Delta)$} 
& 
$\PoinS_{M_{(n_1,m_1,p_1,p_2,m_2,n_2)}^{\text{Dress}}}(t^2)$ \\ \midrule
$(1,0,0,0,0,0,2)$ & $(0,1,0,0,0,0,2)$ & $1+t^2$  \\
$(0,0,0,0,1,0,2)$ & $(0,0,0,0,0,1,2)$ &  \\ \midrule
$(1,1,0,0,0,0,2)$ & $(0,0,0,0,1,1,2)$ & $1+2 t^2+t^4$  \\ \midrule
$(0,0,1,0,0,0,2)$ & & $1+t^2+t^4+t^6$   \\ \midrule
$(0,1,1,0,0,0,2)$ & $(0,0,1,0,1,0,2)$ & $1+2 t^2+2 t^4+2 t^6+t^8$  \\ \midrule
$(1,1,1,0,0,0,2)$ & $(0,1,1,0,1,0,2)$ & $1+3 t^2+4 t^4+4 t^6+3 t^8+t^{10}$ \\
$(0,0,1,0,1,1,2)$ & & \\ \midrule
$(1,1,1,0,1,0,2)$ & $(0,1,1,0,1,1,2)$ & $1+4 t^2+7 t^4+8 t^6+7 t^8+4 
t^{10}+t^{12}$ \\ \midrule
$(1,1,1,0,1,1,2)$  & & $1+5 t^2+11 t^4+15 t^6+15 t^8+11 t^{10}+5 t^{12}+t^{14}$ 
\\ \midrule
$(0,1,1,1,1,0,4)$ & & $1+3 t^2+4 t^4+4 t^6+3 t^8+t^{10}$ \\ \midrule
$(1,1,1,1,1,0,4)$ & $(0,1,1,1,1,1,4)$ & $1+4 t^2+7 t^4+8 t^6+7 t^8+4 
t^{10}+t^{12}$  \\ \midrule
$(1,1,1,1,1,1,4)$ & & $1+5 t^2+11 t^4+15 t^6+15 t^8+11 t^{10}+5 t^{12}+t^{14}$  
\\ \midrule
  \multicolumn{3}{c}{+ $5$ Casimir invariants of degree $4$} \\ \midrule
$(0,1,1,1,0,0,6)$ & $(0,0,1,1,1,0,6)$ & $1+2 t^2+2 t^4+2 t^6+t^8$ \\ \midrule
$(0,0,1,1,1,1,6)$ & $(1,1,1,1,0,0,6)$ & $1+3 t^2+4 t^4+4 t^6+3 t^8+t^{10}$ \\ 
\midrule
$(0,0,1,1,0,0,8)$  & & $1+t^2+t^4+t^6$  \\ \midrule
\multicolumn{3}{c}{+ $1$ Casimir invariant of degree $8$} \\
\bottomrule
\end{tabular}
\caption{All minimal generators for the monoids associated to the cones in the 
fan $F_{\text{Ex.III}}$. Note that we get exactly $24$ generators, which equals 
the number of 
rays in $F_{\text{Ex.III}}$. Therefore, each Hilbert basis $\Hilb(S_\tau)$ 
coincides with the corresponding cone basis of $\tau$.}
\label{tab:min_generators_quiver}
\end{table}
Lastly, we need the classical dressing factors, which are the products of the 
dressing factors for each factor in the quiver gauge group, i.e.
\begin{equation}
 P_{\text{Ex.III}}(t^2;n_1,m_1,p_1,p_2,m_2,n_1)= 
P_{\orm(4)}(t^2;p_1,p_2) \cdot 
\prod_{i=1}^2 P_{\orm(2)}(t^2;n_i) \cdot P_{\sprm(1)}(t^2;m_i) \; .
\end{equation}
The dressing factors for the orthogonal groups are provided 
in~\cite{Cremonesi:2014uva}, and the dressing factors for $\sprm(1)$ are the 
same as for $\su$, see for instance~\cite{Cremonesi:2013lqa}.
We observe that the direct product structure makes it rather easy to deduces 
the dressing behaviour in the relative interior of a cone.

Importing the \texttt{Macaulay2} output into \texttt{Mathematica}, we 
quickly arrive at
\begin{subequations}
\label{eqn:HS_quiver}
\begin{align}
 \HS_{\text{Ex.III}}(t^2)=&\frac{R_{\text{Ex.III}}(t^2)}{\left(1-t^2\right)^6 
\left(1-t^4\right)^5 \left(1-t^8\right)} 
\; ,\\
R_{\text{Ex.III}}(t^2)=&
1
+9 t^2
+49 t^4
+141 t^6
+321 t^8
+505 t^{10}
+709 t^{12}
+754 t^{14} \\
&+709 t^{16}
+505 t^{18}
+321 t^{20}
+141 t^{22}
+49 t^{24}
+9 t^{26}
+t^{28} \; . \notag
\end{align}
\end{subequations}
Let us check the propositions of Sec.~\ref{sec:pole_structure}. 
The rational function~\eqref{eqn:HS_quiver} has a pole of order $12$ at 
$t\to 1$, agreeing with the complex dimension of the Coulomb branch for the 
rank $6$ gauge group.
In addition, the numerator of~\eqref{eqn:HS_quiver} is a palindromic polynomial 
of degree $28$, while the denominator is a polynomial of degree $40$. Thus, the 
difference in degrees is $12$, coinciding with the complex dimension of the 
moduli space. 
 
Next, expanding the Hilbert series~\eqref{eqn:HS_quiver} and the 
corresponding plythestic logarithm, we find
\begin{subequations}
 \begin{align}
  \HS_{\text{Ex.III}}(t^2)&= 1+15 t^2+129 t^4+755 t^6+3462 t^8+13162 
t^{10}+43434 t^{12} 
+\mathcal{O}(t^{14}) \\
 \PL_{\text{Ex.III}}(t^2)&= 15 t^2+9 t^4-60 t^6+177 t^8-176 t^{10}-1494 
t^{12}+\mathcal{O}(t^{14})
 \end{align}
 \end{subequations}
Comparing to the Hilbert basis of Tab.~\ref{tab:min_generators_quiver} and the 
$6$ Casimir invariants of the gauge group $\orm(2)^2 \times \sprm(1)^2 
\times \orm(4)$ of degree $4$, $4$, $4$, $4$, $4$, and $8$, we find that
\begin{enumerate}[(i)]
 \item $(1-t^4)^5 (1-t^8)$ of~\eqref{eqn:HS_quiver} may be chosen to 
correspond to the Casimir invariants of the gauge group.
\item The coefficient $15$ of the $t^2$ in the PL correspond to the fifteen 
degree $2$ minimal generators of Tab.~\ref{tab:min_generators_quiver}. 
\item While the coefficient $9$ of the $t^4$ in the PL correspond to the five 
degree $4$ Casimir invariants plus the four degree $4$ generators of 
Tab.~\ref{tab:min_generators_quiver}. 
\item Thus, we are only missing the four degree $6$ and the one degree $8$ 
generators of Tab.~\ref{tab:min_generators_quiver}, as well as the degree $8$ 
Casimir invariant of the gauge group factor $\orm(4)$. 
 \item The $t^2$ coefficient in expansion of the Hilbert series suggest that 
the global symmetry group of the moduli space is of dimension $15$, which would 
suggest $\sorm(6)$ and $\surm(4)$. Both options are related via $\surm(4) \cong 
\mathrm{Spin}(6)$ and $\mathrm{Spin}(6) \slash \Z_2 \cong \sorm(6)$.
\end{enumerate}
In contrast to the previous two examples, this moduli space is not a nilpotent orbit. For details on those orbits see for instance~\cite{Cabrera:2016vvv}.  
% %%%%%%%%%%%%%%%%%%%%%%%%%%%%%%%%%%%%%%%%%%%%%%%%%%%%%%%%%%%%%%%%%%%%%%%%%%%%%%%%
  \section{Conclusions}
\label{sec:conclusion}
In this article we have explored the implications of understanding the monopole 
formula as being organised by a fan $F$ and its associated collection of 
monoids $S_\tau$, $\tau \in F$. From the mathematical point of view this 
approach is very rich as additive monoids and their associated algebras are 
well-studied. 
Consequently, we have reformulated the entire monopole formula in four ways: 
\begin{enumerate}[(i)]
 \item As twisted sum~\eqref{eqn:rewrite_HS} of Hilbert series for modules 
corresponding to $\Relint(S_\tau)$ times \Poincare\ series for the Casimir 
invariance along $\Relint(S_\tau)$. This form connects the monopole formula 
with two mathematically well-defined objects.
 \item As explicit rational function~\eqref{eqn:rewrite_HS_refinement} by 
utilising a simplicial refinement of the matter fan $F$. Here, the Casimir 
invariants of $\G$ together with the cone generators of $F$ determine 
the denominator. Moreover, this form is particularly useful to prove the pole 
order statements of Prop.~\ref{prop:pole_t=1}-\ref{prop:pole_t=root_of_unity} 
for $t^n = 1$ and $t \to \infty$.
 \item As explicit rational function~\eqref{eqn:rewrite_HS_resolution} by means 
of free resolutions of the lattice ideals for each monoid. The denominator is 
determined by the Casimir invariants of $\G$ and the Hilbert bases of $F$. In 
addition, we found the approach well-suited for the evaluation of the monopole 
formula with computer algebra systems.
 \item As explicit rational function~\eqref{eqn:rewrite_HS_canonical-module} by 
merging triangulations with the properties of canonical modules. Again, the 
Casimir invariants of $\G$ together with the cone generators of $F$ provide the 
denominator. 
\end{enumerate}
The appearing structures and the wealth of known examples strongly suggest that 
the Coulomb branches are Cohen-Macaulay, which is the content of 
Conj.~\ref{conj:Cohen_Macaulay}.

We have provided further insides in the dressing behaviour of monopole 
operators in Prop.~\ref{prop:Casimir_inv}-\ref{prop:Dress_to_faces} and reduced 
the effects for an operator of 
magnetic charge $x$ to the module $M_x^{\text{Dress}}$. The \Poincare\ series 
for this module equals the ratio of dressing factors $P_{\G}(t;x)\slash 
P_{\G}(t;0)$. On the one hand, it demonstrates that only a finite number of 
dressed monopole generators for each $x$ exist. On the other hand, it allows 
to identify a sufficient set of chiral ring generators as in 
Prop.~\ref{prop:chiral_ring}.

Lastly, we have computed the Hilbert series for three quiver gauge theories of 
higher rank using the approach advocated before. In each case, the fan serves 
as ordering scheme to an otherwise very cumbersome task. This demonstrates that 
the novel view point is not limited to the rank $2$ case of our earlier 
work~\cite{Hanany:2016ezz}.

Before closing, we comment on open questions that we would like to address in 
the future. The identification of the chiral ring generators is a first step, 
but we really need to understand the relations between them. There is a precise 
notion of relations/syzygies encoded in the lattice ideal and its resolution. 
However, it is at the moment not clear to us whether they are related to the 
relations on the Coulomb branch itself.
Additionally, it would be interesting to see if the twisted structure of the 
Hilbert series and the local product structure 
$\mathfrak{J}(\hfrak_{\Hh_{x}})^{\Wcal_{\Hh_{x}}} 
\times \K[\Relint(S_\tau) ]$ (for $x \in \Relint(S_\tau)$) allow a global 
understanding of the Coulomb branch.
\section*{Acknowledgements}
We thank Simon Brandhorst, Santiago Cabrera, Bo Feng, Giulia Ferlito, Yang-Hui 
He, Rudolph Kalveks, and Zhenghao Zhong for useful discussions.
A.~H.\ is supported by STFC Consolidated Grant ST\/J0003533\/1, and EPSRC 
Programme Grant EP/K034456/1.
M.~S.\ was supported by the DFG research training group GRK1463 ``Analysis, 
Geometry, and String Theory'' and the Institut für Theoretische Physik of the 
Leibniz Universität Hannover. M.~S.\ is currently supported by Austrian 
Science Fund (FWF) grant P28590.
%%%%%%%%%%%%%%%%%%%%%%%%%%%%%%%%%%%%%%%%%%%%%%%%%%%%%%%%%%%%%%%%%%%%%%%%%%%%%%%
\appendix
\section{Reminder: Algebraic geometry}
\label{app:AlgGeo}
In this appendix we provide the definitions and examples for algebraic concepts 
we have used in the main text. In particular, we elaborate on Cohen-Macaulay and 
Gorenstein rings. Roughly speaking, these are rings with nice 
properties compared to a generic ring and have shown their 
(mathematical and physical) relevance in the study of toric varieties, 
c.f.~\cite{Cox:2011}. For instance, normal toric varietes are Cohen-Macaulay, 
and are in addition Gorenstein if a certain condition on the canonical divisor 
is satisfied.
For this exposition, we follow~\cite{Bruns:Herzog} and refer to 
standard textbooks for a more detailed treatment.

Besides the (Krull) dimension, the \emph{depth} is another important numerical 
invariant of a ring\footnote{To be precise, one considers a Noetherian local 
ring, which we from now on assume.} $R$ or a finite $R$-module $M$. We now 
recall the definition: An element 
$x\in R$ is called \emph{$M$-regular} if $xz=0$ for $z\in M$ implies $z=0$. A 
sequence 
$\boldsymbol{x}= x_1,\ldots,x_n$ of elements in $R$ is called an 
\emph{$M$-regular sequence} if two conditions hold: (i) $x_i$ is an $M \slash 
\langle 
x_1,\ldots,x_{i-1} \rangle$-regular element for $i=1,\ldots,n$, and (ii) 
$M\slash \langle \boldsymbol{x}\rangle \neq 0$.
As example, consider the sequence $z_1,\ldots, z_k$ of variables of 
$R=\K[z_1,\ldots,z_k]$. The first 
condition reads: $z_i$ is regular in $R \slash \langle 
z_1,\ldots,z_{i-1} \rangle \cong \K[z_i,\ldots,z_k]$; while the second 
condition yields the field over which $R$ is defined, i.e.\ $R\slash \langle 
z_1,\ldots,z_k\rangle \cong \K$. Hence, $z_1,\ldots, z_k$  is an $R$-regular 
sequence.

Given an $M$-sequence $\boldsymbol{x}= x_1,\ldots,x_n$, one observes that the 
sequence 
\begin{equation}
 \langle x_1\rangle \subset \langle x_1,x_2\rangle \subset \ldots 
\subset \langle x_1, \ldots,x_n \rangle
\end{equation}
is strictly ascending. An $M$-sequence can be extended to a maximal such 
sequence, i.e.\ an $M$-sequence $\boldsymbol{x}$ is maximal if 
$x_1,\ldots,x_{n+1}$ is not an $M$-sequence for any $x_{n+1}\in R$.  An 
important fact is that all maximal $M$-sequences in an ideal $I$ of $R$ with $I 
M \neq M$ have the same length, called \emph{grade}, provided $M$ is 
finite~\cite[Thm.~1.2.5, p.~10]{Bruns:Herzog}. The depth of $M$ is then defined 
as the 
grade of a maximal ideal $\mathfrak{m}$ of $R$. The relation between depth 
and dimension of an $R$-module $M$ is $\mathrm{depth}(M) \leq \dim(M)$, 
c.f.~\cite[Prop.~1.2.12, p.~12]{Bruns:Herzog}.

With the depth at hand, one defines that a finite $R$-module $M\neq0$ is a 
\emph{Cohen-Macaulay module} if $\mathrm{depth}(M) = \dim(M)$. If $R$ itself is 
a Cohen-Macaulay module, then it is called a \emph{Cohen-Macaulay ring}.
As we know from Lem.~\ref{lem:normal_algebra} and~\ref{lem:canonical_module}, 
a monoid ring $\K[S]$ is Cohen-Macaulay if and only if $S$ is a normal monoid. 

We recall that the \emph{normalisation} of $R$ in the $R$-module $M$ is the 
ring if all elements of $M$ which are integral over $R$. For the special case of 
$M$ being the quotient ring of an integral domain $R$ it is simply called the 
normalisation of $R$. A integral ring is called \emph{normal} if it is own 
normalisation. For monoid rings $\K[S]$, the normality becomes a condition on 
the underlying monoid by Lem.~\ref{lem:normal_algebra}. Let $S$ be a monoid in 
the lattice $\Lambda$, an element $\nu \in \Lambda$ is integral over $S$ if  $c 
\nu \in S$ for some $c\in \NN_{>0}$. The set of all such elements is the 
integral closure $\bar{S}_\Lambda$ of $S$ in $\Lambda$. For a monoid $S$, 
consider 
the group $\Z S$, which is the smallest group containing $S$. 
Then $S$ is called \emph{normal} if it equals its own normalisation 
$\bar{S}\coloneqq \bar{S}_{\Z S}$, i.e.\ $S= \bar{S}$.
Let us consider the following two examples:
\begin{myEx}[non-normal]
Let $S_1\subset \Z^2$ be the monoid generated by 
$\left\{(4,0),(3,1),(1,3),(0,4) \right\}$, see Fig.~\ref{fig:non-normal_monoid}. 
We observe that the point $(2,2)$ does not lie in $S_1$, but in $\Z S_1$. This 
follows from various forms
\begin{equation}
 (2,2) = 2\cdot (3,1) -(4,0) \quad \text{or} \quad (2,2) = 2\cdot (1,3) -(0,4) 
\; .
\end{equation}
By definition of normality, we see that $2\cdot (2,2) \in S_1$ while 
$(2,2)\notin S_1$, implying that $S_1$ is not normal. Therefore, $\K[S_1]$ is 
neither normal nor Cohen-Macaulay. The geometric reason behind is that $S_1$ 
does not originate 
from a strongly convex polyhedral cone. Nevertheless, we can provide the 
explicit description of the monoid ring as  follows
\begin{equation}
 \begin{aligned}
  R_1 &\equiv \K[S_1] \cong \K[z_1,z_2,z_3,z_4] \slash 
  \langle z_2 z_3 - z_1 z_4, z_3^3 - z_2 z_4^2, z_1 z_3^2 - z_2^2 z_4, z_2^3 - 
z_1^2 z_3 \rangle \; , \\
&\for z_1 = x^{(4,0)} \; , z_2 = x^{(3,1)} \; , z_3 = x^{(1,3)} \; , z_4 = 
x^{(0,4)} \; . 
 \end{aligned}
\end{equation}
\end{myEx}
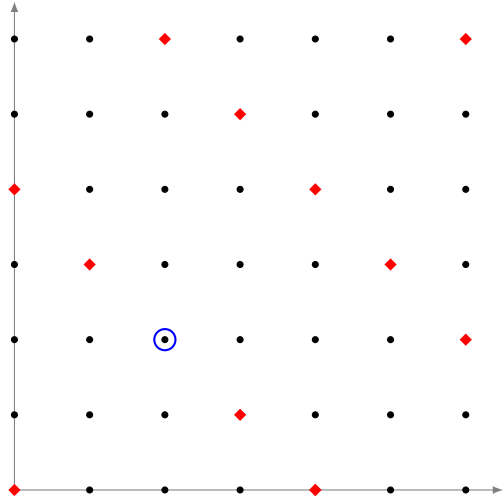
\begin{figure}[h]
\centering
\begin{tikzpicture}
  \coordinate (Origin)   at (0,0);
  \coordinate (XAxisMin) at (0,0);
  \coordinate (XAxisMax) at (6.5,0);
  \coordinate (YAxisMin) at (0,0);
  \coordinate (YAxisMax) at (0,6.5);
  \draw [thin, gray,-latex] (XAxisMin) -- (XAxisMax);%
  \draw [thin, gray,-latex] (YAxisMin) -- (YAxisMax);%
%Draw the monoid
  \foreach \x in {0,1,...,6}{%
      \foreach \y in {0,1,...,6}{%
        \node[draw,circle,inner sep=0.8pt,fill,black] at (\x,\y) {};
	      }
	      }
\draw (0,0) node[diamond,inner sep=1.2pt,fill,red] {};
\draw (4,0) node[diamond,inner sep=1.2pt,fill,red] {};
\draw (3,1) node[diamond,inner sep=1.2pt,fill,red] {};
\draw (1,3) node[diamond,inner sep=1.2pt,fill,red] {};
\draw (0,4) node[diamond,inner sep=1.2pt,fill,red] {};
\draw (6,2) node[diamond,inner sep=1.2pt,fill,red] {};
\draw (5,3) node[diamond,inner sep=1.2pt,fill,red] {};
\draw (4,4) node[diamond,inner sep=1.2pt,fill,red] {};
\draw (4,0) node[diamond,inner sep=1.2pt,fill,red] {};
\draw (2,6) node[diamond,inner sep=1.2pt,fill,red] {};
\draw (3,5) node[diamond,inner sep=1.2pt,fill,red] {};
\draw (6,6) node[diamond,inner sep=1.2pt,fill,red] {};
\draw[thick,blue] (2,2) circle (4pt);
\end{tikzpicture}%
\caption{The monoid $S_1$ spanned by $\left\{(4,0),(3,1),(1,3),(0,4) \right\}$ 
is drawn by red diamonds, while red diamonds and black dots together form 
$\Z^2$. The blue 
circled point is $(2,2)$.  }%
\label{fig:non-normal_monoid}%
\end{figure}%
As a remark, such a non-normal ring would have a peculiar appearance in 
physics, as there exists a certain operator which is not in the chiral ring, 
but any power of it is an element of the chiral ring. Nevertheless, physical 
realisations of non-normal varieties appeared, for example, as Higgs 
branches in~\cite{Hanany:2016gbz}, based on the mathematical results 
of~\cite{kraft:1982}.
\begin{myEx}[normal and Cohen-Macaulay]
Let $S_2\subset \Z^2$ be the the $2$-dimensional monoid generated by 
$\left\{(1,2),(1,1),(1,0),(1,-1) \right\}$, see Fig.~\ref{fig:normal_monoid}. It 
is apparent that $S_2$ is the monoid associated to the cone 
$\cone\{(1,2),(1,-1)\}$ by intersection with the lattice $\Z^2$. Since the cone 
is strongly convex, $S_2$ is normal. Therefore, $\K[S_2]$ is normal as well as 
Cohen-Macaulay. In more detail, the monoid algebra is given by
\begin{equation}
\begin{aligned}
 R_2 &\equiv \K[S_2] \cong \K[z_1,z_2,z_3,z_4]\slash \langle z_3^2 -z_2 z_4, 
z_2 z_3 - z_1 z_4, z_2^2 - z_1 z_3 \rangle \; , \\
&\for z_1=x^{(1,2)} \; , z_2=x^{(1,1)} \; , z_3=x^{(1,0)} \; , z_4=x^{(1,-1)} 
\; .
\end{aligned}
\end{equation}
\end{myEx}
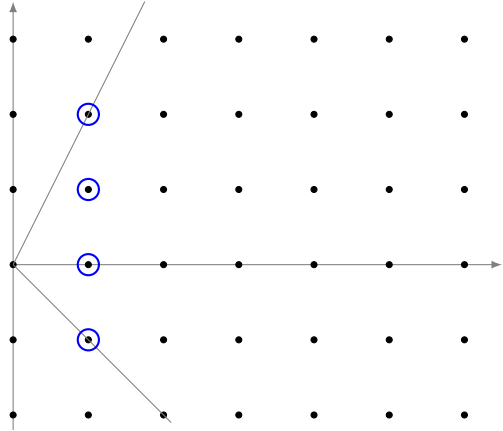
\begin{figure}[h]
\centering
\begin{tikzpicture}
  \coordinate (Origin)   at (0,0);
  \coordinate (XAxisMin) at (0,0);
  \coordinate (XAxisMax) at (6.5,0);
  \coordinate (YAxisMin) at (0,-2.2);
  \coordinate (YAxisMax) at (0,3.5);
  \draw [thin, gray,-latex] (XAxisMin) -- (XAxisMax);%
  \draw [thin, gray,-latex] (YAxisMin) -- (YAxisMax);%
%Draw the monoid
  \foreach \x in {0,1,...,6}{%
      \foreach \y in {-2,-1,...,3}{%
        \node[draw,circle,inner sep=0.8pt,fill,black] at (\x,\y) {};
	      }
	      }
  \draw [thin, gray] (Origin) -- (7/4*1,7/4*2);%
  \draw [thin, gray] (Origin) -- (2.1*1,-2.1*1);%
\draw[thick,blue] (1,2) circle (4pt);
\draw[thick,blue] (1,1) circle (4pt);
\draw[thick,blue] (1,0) circle (4pt);
\draw[thick,blue] (1,-1) circle (4pt);
\end{tikzpicture}%
\caption{The monoid $S_2$ spanned by $\left\{(1,2),(1,1),(1,0),(1,-1) \right\}$ 
and the blue circled points are the elements of the Hilbert basis.}%
\label{fig:normal_monoid}%
\end{figure}%

Of fundamental importance for the study of a Cohen-Macaulay ring $R$ is the 
canonical module $\omega_R$ of it. We refrain from the generic 
definition~\cite[Def.~3.3.1, p.~107]{Bruns:Herzog} and settle on 
the statements valid for monoid algebras $R=\K[S]$. In general, the existence 
and uniqueness of $\omega_R$ has to be addressed; however, for normal monoid 
rings the canonical module exists and equals the unique ideal spanned by 
the monomials $x^\nu$ with $\nu \in \Relint(S)$.
In the simplest case, we consider the monoid $S$ spanned by $\{(1,0),(0,1)\}$, 
for which $\K[S] \cong \K[z_1,z_2]$ for $z_1=x^{(1,0)}$ and $z_2=x^{(0,1)}$. 
Then $\Relint(S)=\{(\nu_1,\nu_2)\in \NN^2 |\,\nu_1>0,\, \nu_2>0 \}$ such that 
$\omega_R=\langle z_1 \cdot z_2\rangle$.

An interesting sub-class of Cohen-Macaulay rings are so-called \emph{Gorenstein 
rings}. The defining property is that the ring is (isomorphic to) its own 
canonical module. 
For a monoid ring this translates into the condition: there exists a $\nu 
\in \Relint(S)$ such that $\Relint(S)=\nu + S$. Let us consider the following 
example:
\begin{myEx}[Gorenstein]
Let $S_3\subset \Z^2$ be the monoid spanned by $\{(1,1),(1,0),(1,-1)\}$, as 
depcited in Fig.~\ref{fig:Gorenstein_monoid}. It is convenient represent points 
in the two sub-monoids $S_3 \cap \{\pm\nu_2\geq0\}$ via
\begin{subequations}
\begin{align}
 (\nu_1,\nu_2) &= (\nu_1-\nu_2) \cdot (1,0) + \nu_2 \cdot (1,1) \for \nu_1\geq 
\nu_2 \geq 0 \; , \\
(\nu_1,-\nu_2) &= (\nu_1-\nu_2) \cdot (1,0) + \nu_2 \cdot (1,-1) \for \nu_1\geq 
\nu_2 \geq 0 \; .
\end{align}
\end{subequations}
In other words, the monoid $S_3$ is sort of foliated along the $1$-dimensional 
monoid 
spanned by $(1,0)$, which then implies that 
\begin{equation}
 \Relint(S_3) = (1,0) + S_3 \; .
\end{equation}
Therefore, the monoid algebra $\K[S_3]$ is not only Cohen-Macaulay, but also 
Gorenstein.  In detail
\begin{subequations}
\begin{align}
 R_3&\equiv \K[S_3] \cong \K[z_1,z_2,z_3]\slash \langle z_1 \cdot z_2 -z_3^2 
\rangle \und \omega_{R_3} = \langle z_3\rangle \equiv z_3 \K[S_3]   \; ,\\
&\for z_1 = x^{(1,1)}\; , z_2 = x^{(1,-1)} \; , z_3 = 
x^{(1,0)} \; . 
\end{align}
\end{subequations}
Thus, the canonical module is just a translate of $R_3$, i.e.\ all 
elements are shifted in degree by $(1,0)$.
\end{myEx}
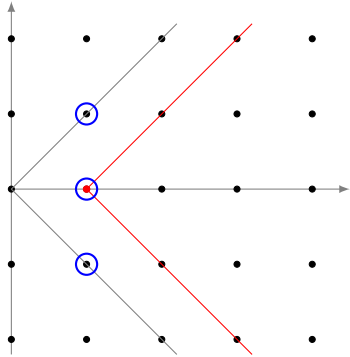
\begin{figure}[h]
\centering
\begin{tikzpicture}
  \coordinate (Origin)   at (0,0);
  \coordinate (XAxisMin) at (0,0);
  \coordinate (XAxisMax) at (4.5,0);
  \coordinate (YAxisMin) at (0,-2.2);
  \coordinate (YAxisMax) at (0,2.5);
  \draw [thin, gray,-latex] (XAxisMin) -- (XAxisMax);%
  \draw [thin, gray,-latex] (YAxisMin) -- (YAxisMax);%
%Draw the monoid
  \foreach \x in {0,1,...,4}{%
      \foreach \y in {-2,-1,...,2}{%
        \node[draw,circle,inner sep=0.8pt,fill,black] at (\x,\y) {};
	      }
	      }
  \draw [thin, gray] (Origin) -- (2.2,2.2);%
  \draw [thin, gray] (Origin) -- (2.2,-2.2);%
\draw (1,0) node[circle,inner sep=1pt,fill,red] {};
  \draw [thin, red] (1,0) -- (1+2.2,2.2);%
  \draw [thin, red] (1,0) -- (1+2.2,-2.2);%
\draw[thick,blue] (1,1) circle (4pt);
\draw[thick,blue] (1,0) circle (4pt);
\draw[thick,blue] (1,-1) circle (4pt);
\end{tikzpicture}%
\caption{The monoid $S_3$ spanned by $\left\{(1,1),(1,0),(1,-1) \right\}$ 
and the blue circled points are the elements of the Hilbert basis. The red 
point $(1,0)$ is a point in the relative interior such that the shift of 
$S_3$ by $(1,0)$ (denoted by the red lines) equals the relative interior.}%
\label{fig:Gorenstein_monoid}%
\end{figure}%
% 
% \clearpage
% 
%%%%%%%%%%%%%%%%%%%%%%%%%%%%%%%%%%%%%%%%%%%%%%%%%%%%%%%%%%%%%%%%%%%%%%%%%%
% 
 \bibliographystyle{JHEP}     % Zitierstil: alpha = [Nam88]
 {\footnotesize{\bibliography{references}}}

\providecommand{\href}[2]{#2}\begingroup\raggedright\begin{thebibliography}{10}

\bibitem{Cremonesi:2013lqa}
S.~Cremonesi, A.~Hanany, and A.~Zaffaroni, {\it {Monopole operators and Hilbert
  series of Coulomb branches of $3d$ $\mathcal{N} = 4$ gauge theories}},  {\em
  JHEP} {\bf 01} (2014) 005, [\href{http://arxiv.org/abs/1309.2657}{{\tt
  arXiv:1309.2657}}].

\bibitem{Cremonesi:2014vla}
S.~Cremonesi, A.~Hanany, N.~Mekareeya, and A.~Zaffaroni, {\it {Coulomb branch
  Hilbert series and Three Dimensional Sicilian Theories}},  {\em JHEP} {\bf
  09} (2014) 185, [\href{http://arxiv.org/abs/1403.2384}{{\tt
  arXiv:1403.2384}}].

\bibitem{Cremonesi:2014kwa}
S.~Cremonesi, A.~Hanany, N.~Mekareeya, and A.~Zaffaroni, {\it {Coulomb branch
  Hilbert series and Hall-Littlewood polynomials}},  {\em JHEP} {\bf 09} (2014)
  178, [\href{http://arxiv.org/abs/1403.0585}{{\tt arXiv:1403.0585}}].

\bibitem{Cremonesi:2014xha}
S.~Cremonesi, G.~Ferlito, A.~Hanany, and N.~Mekareeya, {\it {Coulomb Branch and
  The Moduli Space of Instantons}},  {\em JHEP} {\bf 12} (2014) 103,
  [\href{http://arxiv.org/abs/1408.6835}{{\tt arXiv:1408.6835}}].

\bibitem{Cremonesi:2014uva}
S.~Cremonesi, A.~Hanany, N.~Mekareeya, and A.~Zaffaroni, {\it
  {T$_{\rho}^{\sigma}$ (G) theories and their Hilbert series}},  {\em JHEP}
  {\bf 01} (2015) 150, [\href{http://arxiv.org/abs/1410.1548}{{\tt
  arXiv:1410.1548}}].

\bibitem{Hanany:2016ezz}
A.~Hanany and M.~Sperling, {\it {Coulomb branches for rank 2 gauge groups in 3d
  $ \mathcal{N}=4 $ gauge theories}},  {\em JHEP} {\bf 08} (2016) 016,
  [\href{http://arxiv.org/abs/1605.00010}{{\tt arXiv:1605.00010}}].

\bibitem{Hanany:2015via}
A.~Hanany, C.~Hwang, H.~Kim, J.~Park, and R.-K. Seong, {\it {Hilbert Series for
  Theories with Aharony Duals}},  {\em JHEP} {\bf 11} (2015) 132,
  [\href{http://arxiv.org/abs/1505.02160}{{\tt arXiv:1505.02160}}]. [Addendum:
  JHEP04,064(2016)].

\bibitem{Cremonesi:2015dja}
S.~Cremonesi, {\it {The Hilbert series of 3d $\mathcal{N}=2$ Yang–Mills
  theories with vectorlike matter}},  {\em J. Phys.} {\bf A48} (2015), no.~45
  455401, [\href{http://arxiv.org/abs/1505.02409}{{\tt arXiv:1505.02409}}].

\bibitem{Cremonesi:2016nbo}
S.~Cremonesi, N.~Mekareeya, and A.~Zaffaroni, {\it {The moduli spaces of $3d$
  $\mathcal{N} \ge 2$ Chern-Simons gauge theories and their Hilbert series}},
  \href{http://arxiv.org/abs/1607.05728}{{\tt arXiv:1607.05728}}.

\bibitem{Carta:2016fjb}
F.~Carta and H.~Hayashi, {\it {Hilbert series and mixed branches of $T[SU(N)]$
  theory}},  \href{http://arxiv.org/abs/1609.08034}{{\tt arXiv:1609.08034}}.

\bibitem{Nakajima:2015txa}
H.~Nakajima, {\it {Towards a mathematical definition of Coulomb branches of
  $3$-dimensional $\mathcal N=4$ gauge theories, I}},
  \href{http://arxiv.org/abs/1503.03676}{{\tt arXiv:1503.03676}}.

\bibitem{Nakajima:2015gxa}
H.~Nakajima, {\it {Questions on provisional Coulomb branches of $3$-dimensional
  $\mathcal N=4$ gauge theories}},  \href{http://arxiv.org/abs/1510.03908}{{\tt
  arXiv:1510.03908}}.

\bibitem{Braverman:2016wma}
A.~Braverman, M.~Finkelberg, and H.~Nakajima, {\it {Towards a mathematical
  definition of Coulomb branches of $3$-dimensional $\mathcal N=4$ gauge
  theories, II}},  \href{http://arxiv.org/abs/1601.03586}{{\tt
  arXiv:1601.03586}}.

\bibitem{Bullimore:2015lsa}
M.~Bullimore, T.~Dimofte, and D.~Gaiotto, {\it {The Coulomb Branch of 3d
  $\mathcal{N}=4$ Theories}},  \href{http://arxiv.org/abs/1503.04817}{{\tt
  arXiv:1503.04817}}.

\bibitem{Bullimore:2016hdc}
M.~Bullimore, T.~Dimofte, D.~Gaiotto, J.~Hilburn, and H.-C. Kim, {\it {Vortices
  and Vermas}},  \href{http://arxiv.org/abs/1609.04406}{{\tt
  arXiv:1609.04406}}.

\bibitem{Goddard:1976qe}
P.~Goddard, J.~Nuyts, and D.~I. Olive, {\it {Gauge Theories and Magnetic
  Charge}},  {\em Nucl. Phys.} {\bf B125} (1977) 1.

\bibitem{Englert:1976ng}
F.~Englert and P.~Windey, {\it {Quantization Condition for 't Hooft Monopoles
  in Compact Simple Lie Groups}},  {\em Phys. Rev.} {\bf D14} (1976) 2728.

\bibitem{tHooft:1977hy}
G.~'t~Hooft, {\it {On the Phase Transition Towards Permanent Quark
  Confinement}},  {\em Nucl. Phys.} {\bf B138} (1978) 1.

\bibitem{Borokhov:2002cg}
V.~Borokhov, A.~Kapustin, and X.-k. Wu, {\it {Monopole operators and mirror
  symmetry in three-dimensions}},  {\em JHEP} {\bf 12} (2002) 044,
  [\href{http://arxiv.org/abs/hep-th/0207074}{{\tt hep-th/0207074}}].

\bibitem{Borokhov:2002ib}
V.~Borokhov, A.~Kapustin, and X.-k. Wu, {\it {Topological disorder operators in
  three-dimensional conformal field theory}},  {\em JHEP} {\bf 11} (2002) 049,
  [\href{http://arxiv.org/abs/hep-th/0206054}{{\tt hep-th/0206054}}].

\bibitem{Gaiotto:2008ak}
D.~Gaiotto and E.~Witten, {\it {S-Duality of Boundary Conditions In N=4 Super
  Yang-Mills Theory}},  {\em Adv. Theor. Math. Phys.} {\bf 13} (2009), no.~3
  721--896, [\href{http://arxiv.org/abs/0807.3720}{{\tt arXiv:0807.3720}}].

\bibitem{Benna:2009xd}
M.~K. Benna, I.~R. Klebanov, and T.~Klose, {\it {Charges of Monopole Operators
  in Chern-Simons Yang-Mills Theory}},  {\em JHEP} {\bf 01} (2010) 110,
  [\href{http://arxiv.org/abs/0906.3008}{{\tt arXiv:0906.3008}}].

\bibitem{Bashkirov:2010kz}
D.~Bashkirov and A.~Kapustin, {\it {Supersymmetry enhancement by monopole
  operators}},  {\em JHEP} {\bf 05} (2011) 015,
  [\href{http://arxiv.org/abs/1007.4861}{{\tt arXiv:1007.4861}}].

\bibitem{Ziegler:1995}
G.~M. Ziegler, {\em {Lectures on Polytopes}}, vol.~152 of {\em {Graduate Texts
  in Mathematics}}.
\newblock Springer New York, 1995.
\newblock {Updated Seventh Printing of the First Edition}.

\bibitem{Bruns:Herzog}
W.~Bruns and H.~J. Herzog, {\em Cohen-Macaulay Rings}.
\newblock Cambridge University Press, 1998.

\bibitem{Koch:2003}
R.~Koch, {\em Affine Monoids, Hilbert Bases and Hilbert Functions}.
\newblock PhD thesis, Universität Osnabrück, 2003.

\bibitem{Bruns:2009}
W.~Bruns and J.~Gubeladze, {\em Polytopes, rings, and {$K$}-theory}.
\newblock Springer Monographs in Mathematics. Springer, Dordrecht, 2009.

\bibitem{Miller:2005}
E.~Miller and B.~Sturmfels, {\em {Combinatorial Commutative Algebra}}, vol.~227
  of {\em {Graduate Texts in Mathematics}}.
\newblock Springer New York, 2005.

\bibitem{Sebo:1990}
A.~Seb\"{o}, {\it Hilbert bases, caratheodory's theorem and combinatorial
  optimization},  in {\em Proceedings of the 1st Integer Programming and
  Combinatorial Optimization Conference}, (Waterloo, Ont., Canada, Canada),
  pp.~431--455, University of Waterloo Press, 1990.

\bibitem{Cox:Little}
D.~Cox, J.~Little, and D.~O'Shea, {\em Using algebraic geometry}, vol.~185 of
  {\em Graduate Texts in Mathematics}.
\newblock Springer-Verlag, New York, 1998.

\bibitem{Stanley:1978}
R.~P. Stanley, {\it Hilbert functions of graded algebras},  {\em Advances in
  Math.} {\bf 28} (1978), no.~1 57--83.

\bibitem{Ferlito:2016grh}
G.~Ferlito and A.~Hanany, {\it {A tale of two cones: the Higgs Branch of Sp(n)
  theories with 2n flavours}},  \href{http://arxiv.org/abs/1609.06724}{{\tt
  arXiv:1609.06724}}.

\bibitem{Macaulay2book}
D.~Eisenbud, D.~R. Grayson, M.~E. Stillman, and B.~Sturmfels, {\em
  {Computations in algebraic geometry with Macaulay 2}}, vol.~8 of {\em
  {Algorithms and Computations in Mathematics}}.
\newblock Springer-Verlag, 2001.

\bibitem{Polyhedra:Birkner}
R.~Birkner, {\it Polyhedra: A package for computations with convex polyhedral
  objects},  {\em Journal of Software for Algebra and Geometry} {\bf 1} (2009),
  no.~1 11--15.

\bibitem{Benvenuti:2010pq}
S.~Benvenuti, A.~Hanany, and N.~Mekareeya, {\it {The Hilbert Series of the One
  Instanton Moduli Space}},  {\em JHEP} {\bf 06} (2010) 100,
  [\href{http://arxiv.org/abs/1005.3026}{{\tt arXiv:1005.3026}}].

\bibitem{Cabrera:2016vvv}
S.~Cabrera and A.~Hanany, {\it {Branes and the Kraft-Procesi Transition}},
  \href{http://arxiv.org/abs/1609.07798}{{\tt arXiv:1609.07798}}.

\bibitem{Cox:2011}
D.~Cox, J.~Little, and H.~Schenck, {\em {Toric Varieties}}, vol.~124 of {\em
  {Graduate Studies in Mathematics}}.
\newblock American Mathematical Soc., 2011.

\bibitem{Hanany:2016gbz}
A.~Hanany and R.~Kalveks, {\it {Quiver Theories for Moduli Spaces of Classical
  Group Nilpotent Orbits}},  {\em JHEP} {\bf 06} (2016) 130,
  [\href{http://arxiv.org/abs/1601.04020}{{\tt arXiv:1601.04020}}].

\bibitem{kraft:1982}
H.~Kraft and C.~Procesi, {\it On the geometry of conjugacy classes in classical
  groups},  {\em Commentarii Mathematici Helvetici} {\bf 57} (1982), no.~1
  539--602.

\end{thebibliography}\endgroup

\end{document}